\documentclass[aps,preprint,pre,superscriptaddress]{revtex4-1}
\usepackage{graphicx,subfigure,tabularx}
\usepackage[dvipsnames]{xcolor}

\newcolumntype{Y}{>{\centering\arraybackslash}X}

\begin{document}
\raggedbottom % sometimes need this to avoid latex putting in inconsistent skips between paragraphs

\title{Universal Features in Phonological Neighbor Networks}
\date{\today}

\author{Kevin S. Brown}
\affiliation{Department of Biomedical Engineering, University of Connecticut, Storrs, CT, USA}
\affiliation{Department of Physics, University of Connecticut, Storrs, CT, USA}
\affiliation{Institute for Systems Genomics, University of Connecticut, Storrs, CT, USA}
\affiliation{Connecticut Institute for the Brain \& Cognitive Sciences, University of Connecticut, Storrs, CT, USA}
\email{kevin.s.brown@uconn.edu}
\homepage{http://kbrown.research.uconn.edu}
\author{Paul D. Allopenna}
\affiliation{Department of Psychology, University of Connecticut, Storrs, CT, USA}
\author{William R. Hunt}
\affiliation{Department of Biomedical Engineering, University of Connecticut, Storrs, CT, USA}
\author{Rachael Steiner}
\affiliation{Department of Psychology, University of Connecticut, Storrs, CT, USA}
\author{Elliot Saltzman}
\affiliation{Department of Physical Therapy and Athletic Training, Boston University, Boston, MA, USA}
\author{Ken McRae}
\affiliation{Department of Psychology, University of Western Ontario, London, Ontario, CAN}
\affiliation{Brain \& Mind Institute, University of Western Ontario, London, Ontario, CAN}
\author{James S. Magnuson}
\affiliation{Connecticut Institute for the Brain \& Cognitive Sciences, University of Connecticut, Storrs, CT, USA}
\affiliation{Department of Psychology, University of Connecticut, Storrs, CT, USA}


\begin{abstract}
Human speech perception involves transforming a countinous acoustic signal into discrete linguistically meaningful units, such as phonemes, while simultaneously
causing a listener to activate words that are similar to the spoken utterance and to each other.  The Neighborhood Activation Model
(NAM~\cite{Luce:1986,Luce:1998}) posits that phonological neighbors (two forms [words] that differ by one phoneme) compete significantly for recognition as a spoken word is
heard. This definition of phonological similarity can be extended to an entire corpus of forms to produce a phonological neighbor
network~\cite{Vitevitch:2008} (PNN). We study PNNs for five languages: English, Spanish, French, Dutch, and German. Consistent with previous work,
we find that the PNNs share a consistent set of topological features.  Using an approach that
generates random lexicons with increasing levels of phonological realism, we show that even random forms with minimal relationship to any real language, combined with only
the empirical distribution of language-specific phonological form lengths, are sufficient to produce the topological properties observed in the real language PNNs.  The
resulting pseudo-PNNs are insensitive to the level of lingusitic realism in the random lexicons but quite sensitive to the shape of the form length distribution. We therefore
conclude that ``universal'' features seen across multiple languages are really string universals, not language universals, and arise primarily due to limitations in the
kinds of networks generated by the one-step neighbor definition.  Taken together, our results indicate that caution is warranted when linking the dynamics of
human spoken word recognition to the topological properties of PNNs, and that the investigation of alternative similarity metrics for phonological forms should be a priority.
\end{abstract}

\maketitle

\section{\label{sec:intro}Introduction}

The preception and recognition of acoustic speech, known in psycholinguistics as \textit{spoken word recognition} (SWR), requires that human listeners rapidly map highly
variable acoustic signals onto stable linguistically relevant categories (in this case, phonemes, i.e.~the consonants and vowels that comprise a language's basic
sound inventory) and then piece together sequences of phonemes into words, all without robust cues to either phoneme or word boundaries (see here~\cite{Fowler:2012,
Magnuson:2013} for reviews). Decades of research on human spoken word recognition has led to a consensus on three broad
principles: (1) SWR occurs in a continuous and incremental fashion as a spoken target word unfolds over time, (2) words in memory are activated proportionally to their
similarity with the acoustic signal as well their prior probability (computed as a function of their frequency of occurrence) in the language, and (3) activated words compete
for recognition. A key difference between theories is how to characterize signal-to-word and word-to-word similarity. Most theories incorporate some set some sort of
similarity threshold, and pairs of words meeting that threshold are predicted to strongly activate each other and compete. Perhaps the most influential definition for the
phonological similarity of spoken words is the concept of phonological neighbors posited under the Neighborhood Activation Model (NAM) by Luce and
colleagues~\cite{Luce:1986,Luce:1998}. NAM includes a gradient similarity metric and a threshold metric, although only the latter is widely used (and we focus on it here).
The threshold metric defines neighbors based on the Deletion-Addition-Substitution (DAS) string metric, which states that two words are neighbors (i.e., they are sufficiently
similar to strongly activate one another and compete) if they differ by no more than the deletion, addition, or substitution of a single phoneme. Thus, \textit{cat} has the
deletion neighbor \textit{at}, addition neighbors \textit{scat} and \textit{cast}, and many substitution neighbors, such as \textit{bad}, \textit{cot}, and \textit{can}.
NAM predicts that a target word's recognizability is determined according to a simple \textit{frequency-weighted neighborhood probability rule} which is defined by the
ratio of the target word's prior probability to the summed prior probability of all its DAS-linked neighbors. The NAM rule predicts a greater proportion of the variance
in spoken word recognition latencies (10-27\%, depending on task [lexical decision, naming, or identification in noise] and conditions [signal-to-noise ratio]~\cite{Luce:1986})
than any other measure that has been tested (e.g.,~log word frequency alone accounted for 5-10\% of variance in Luce's studies).

The focus of the NAM approach has typically been used to characterize the recognizability of single words according to the sizes (densities) of their locally defined
neighborhoods.  More recently, it has been realized that viewing the structure of the phonological lexicaon globally as a complex network enables the probing of
connections between both large and small scale network topology and human spoken word recognition.  Thus, rather than considering a word and its neighbors in isolation,
the set of neighbor relationships for an entire lexicon can be represented as an unweighted, undirected graph~\cite{Vitevitch:2008} in which words (phonological forms)
are represented by nodes and two words are joined by an edge if they meet the standard NAM DAS threshold.  The NAM approach can be translated to the network context to
mean that (frequency-weighted) node degree is important for predicting latencies in spoken word recognition.  There are also prior indications that other topological
properties (e.g.~node clustering coefficient~\cite{Chan:2009,Chan:2010}, closeness centrality~\cite{Iyengar:2012}, and second neighbor density~\cite{Siew:2017}) may
also explain some aspects of SWR that the frequency-weighted neighborhood probability it is based upon does not.

Previous studies have shown that what we will call the \textit{phonological neighbor network}, or PNN, for English has some features of both
Watts-Strogatz~\cite{Watts:1998} and Barabasi-Albert~\cite{Barabasi:1999} graphs. It has a relatively short mean geodesic path length and high clustering coefficient,
but also has a degree distribution that is at least partially power law~\cite{Vitevitch:2008}. Subsequent analyses of additional
languages (English, Spanish, Hawaiian, Basque, and Mandarin) have shown these characteristics to be broadly shared across languages when PNN graphs are constructed
using NAM's DAS rule~\cite{Arbesman:2010a}. On the basis of these results, Vitevich and colleagues have assigned importance to these
language ``universals'' and argued that many of these properties are sensible if not essential (e.g.~, high degree assortativity, which measures the tendency
of nodes to be connected to other nodes of similar degree, can buffer against network damage)~\cite{Arbesman:2010a}.

However, making claims about SWR on the basis of the properties of PNNs alone is potentially fraught for at least two reasons. First, PNNs are static representations of
lexical structure, whereas spoken words are processed incrementally over time. Second, different measures of word similarity will result in radically different PNNs.
NAM's DAS rule is based on a relatively simple string distance metric that provides a local measure of inter-word similarity that is insensitive to the sequence of
phonemes in a word.  Thus, while NAM's DAS metric accounts for substantial variance using a regression-based approach (predicting response latencies for many words),
there is substantial evidence from studies examining competition between specific pairs of words with different patterns of position-dependent phonological overlap that
words whose onsets overlap compete more strongly than words that are matched in DAS similarity but whose onsets are mismatched (e.g., \textit{battle} would compete more
strongly with \textit{batter} than with \textit{cattle}~\cite{Allopenna:1998}).  Marslen-Wilson and colleagues~\cite{MarslenWilson:1978,MarslenWilson:1993}
proposed a threshold metric that gives primacy to onset similarity. They focused on the notion (consistent with many priming
and gating studies~\cite{MarslenWilson:1993}) that the ``cohort'' of words activated by a spoken word is restricted to words overlapping in their first
two phonemes. Thus, the cohort competitors of \textit{cat} include not just DAS neighbors overlapping at onset (\textit{can}, \textit{cab}, \textit{cast}) but also
longer words that would not be DAS neighbors (\textit{cattle}, \textit{castle}, \textit{cabinet}). In addition, the cohort metric predicts that rhyme (i.e.~a word's
vowel and following consonants) neighbors (\textit{cat-bat}, \textit{cattle-battle}) do not compete because they mismatch at onset, despite high DAS similarity.
A PNN based on a simple onset cohort rule (connect words that overlap in the first two phonemes) would obviously have very different structure than a DAS-based PNN.
When using PNNs to compare lexical structure between languages, we must consider the potential role of the similarity metric itself in determining the network's
structure and topology. This possibility calls into question any universal (language-independent) claims about SWR based on DAS networks. Prior work has demonstrated
that this is likely true at least in English, as PNNs constructed from a random lexicon with the same phonological constraints as English are basically indistinguishable
from the real language network~\cite{GP:2009,Stella:2015}.

Here, we explore this possibility further by extending DAS-based PNNs to four languages in addition to English: Spanish, French, German, and Dutch. We show
that PNNs for these languages have degree distributions and topological properties similar to PNNs previously constructed for English, Spanish, Hawaiian,
Mandarin, and Basque~\cite{Arbesman:2010a}.  We then show, by separating words by number of syllables, that all five language
networks consist of aggregations of at least two very different networks, as has been previously suggested for English~\cite{GP:2009}.  We also note for the first time
the effects of homophones such as \textit{bare} and \textit{bear} on PNN structure.  Finally, using a set of models that generate random lexicons with varying levels of
phonological realism, we show that even extremely simple random lexicons, along with language-specific phoneme inventories and distributions of phonological form length
(that is, the frequency distributions of words of different lengths, ignoring all other linguistic details), can create pseudo-PNNs that share all the properties of real PNNs.
 In fact, adding phonological constraints (e.g., phonotactic constraints on phoneme sequences) does very little to improve pseudo-PNN match to language-based PNNs.  While
 these pseudo-PNNs are quite insensitive to the level of realism in the lexicon, they are extremely sensitive to the empirical form length distribution, which we show drives
all of the observed differences among English, French, German, Dutch, and Spanish.  Our results suggest that the primary determinant of the observed topology of PNNs
is the neighbor definition itself, which dramatically limits the network structures possible in PNNs. In addition, our work strongly motivates the consideration of alternate
phonological similarity metrics and suggests that it is important to try to understand the formative dynamics underlying the observed phonological form length distributions.

\section{\label{sec:data}Data}

We used the freely available online CLEARPOND~\cite{Marian:2012} database to construct DAS-based PNNs for five languages: English, Dutch, German, French, and Spanish.
CLEARPOND is described in detail elsewhere~\cite{Marian:2012}, but in brief, it includes phonological transcriptions of orthographic forms and frequency information for
over $27,000$ words from each language. Frequency information for English~\cite{Brysbaert:2009}, Dutch~\cite{Keuleers:2010}, German\cite{Brysbaert:2011}, and
Spanish~\cite{Cuetos:2011} is derived from the SUBTLEX database which counts word occurrences in television and movie subtitles. French frequency information is derived
from Lexique~\cite{New:2004}, a fusion of an older French language database (Frantext) with word occurrence information derived from webpages. For all five languages we
constructed PNNs based on the DAS rule described above: two words were neighbors and therefore linked with a bidirectional, unweighted edge, if they differed by no more
than a single phoneme deletion, addition, or substitution. After PNN construction, we found that, in each language, a significant percentage of the words had no
phonological neighbors, ranging from $24\%$ (French) to $45\%$ (Dutch). All singleton words were excluded from any further analysis, since their topological properties
are either trivial (e.g.~they are all degree zero) or undefined (e.g.~clustering coefficient). In all five languages, the mean length of the neighborless words is larger
than that of the words with neighbors, but this difference is not statistically significant (permutation test).

\section{\label{sec:analysis}Empirical Analysis of Phonological Neighbor Networks}

\subsection{\label{sec:dd}Degree Distributions and Topology}

\begin{figure}[ht!]
	\begin{center}
		\includegraphics[width=\columnwidth]{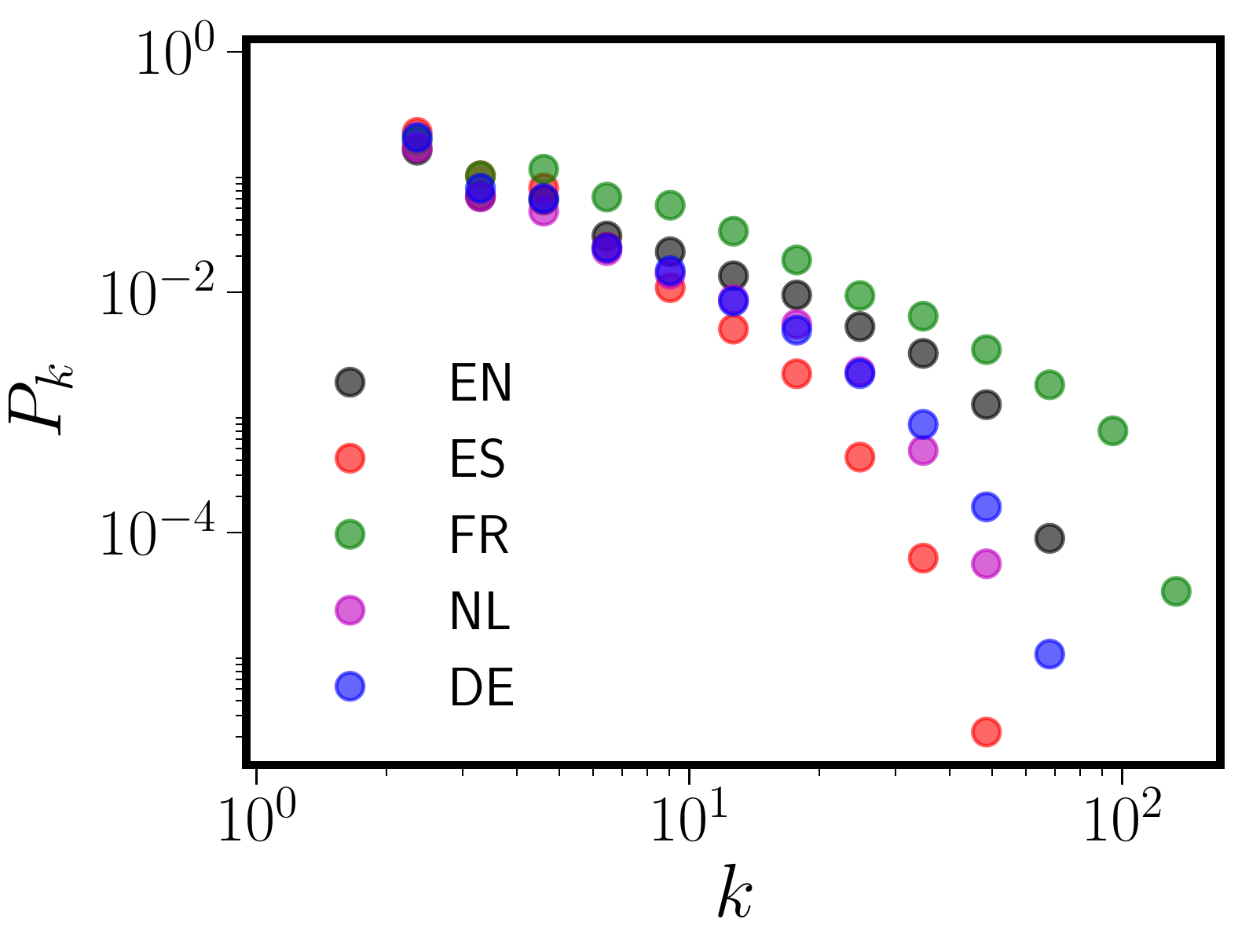}
	\end{center}
	\caption{Log-log plot of the logarithmically binned phonological neighbor network (PNN) degree distribution $P_k$ for five languages:
	English (EN), Spanish (ES), French (FR), Dutch (NL), and German (DE).}
	\label{fig:degreedist}
\end{figure}

Figure~\ref{fig:degreedist} shows the degree distributions for the five PNNs constructed from the CLEARPOND data (compare
also to Figure 10 in the original CLEARPOND paper~\cite{Marian:2012}), and Table~\ref{tab:topology} gives a summary of some of the common topological measures
employed in the empirical analysis of networks, all of which have been specifically highlighted in prior PNN research.  All five language
degree distributions are best fit (via maximum likelihood) by a truncated power law, as tested via likelihood ratio~\cite{Clauset:2009}. In addition, we observe that
all PNNs have: (\textit{i}) relatively high clustering, (\textit{ii}) short mean geodesic paths, (\textit{iii}) extraordinarily high values of
degree assortativity, and (\textit{iv}) relatively small giant connected components (the largest connected subgraph in the network). Thus, all five PNNs have similar
degree distributions and topological characteristics, and they combine some features of Watts-Strogatz~\cite{Watts:1998} graphs (high clustering) with Barabasi-Albert
graphs~\cite{Barabasi:1999} (power law degree distribution). High degree assortativity and small giant component sizes are features of the PNNs that are not displayed
by either WS or BA graphs. These features are all consistent with previous studies on English alone~\cite{Vitevitch:2008,GP:2009} and other languages not studied
here~\cite{Arbesman:2010a}.

The grouping of languages in Figure~\ref{fig:degreedist} is rather surprising. Essentially, Spanish is by itself, Dutch and German have quite
similar degree distributions, and English and French are grouped together. One might expect different clustering based on language typology; for example,
with the two Romance languages (French and Spanish) grouped together.  We will show that the observed clustering can be explained without any reference to the specific
history of words.  Instead, the structure of the phonological form length distribution, along with target language phoneme frequencies, are all that is required.

\begin{table}[h]
	\begin{center}
		\renewcommand{\tabcolsep}{0.20cm}
		\begin{tabular}{c|ccccc}
 			& \textbf{EN} & \textbf{NL} & \textbf{DE} & \textbf{ES} & \textbf{FR} \\
			\hline
			$N$ & 18983 & 15360 & 17227 & 20728 & 21177 \\
			$m$ & 76092 & 36158 & 41970 & 36111 & 145426 \\
			$\bar{k}$ & 8.01 & 4.71 & 4.87 & 3.48 & 13.7 \\
			\textbf{GC size} & 0.66 & 0.56 & 0.58 & 0.43 & 0.74 \\
			$C$ (all/GC) & 0.23/0.28 & 0.16/0.23 & 0.21/0.24 & 0.18/0.20 & 0.24/0.25 \\
			$l$ & 6.68 & 8.48 & 8.73 & 9.41 & 6.85 \\
			$\alpha$ & 1.0$^{*}$ & 1.84$^{*}$ & 1.2$^{*}$ & 2.1$^{*}$ & 1.04$^{*}$ \\
			$r$ (all/GC) & 0.73/0.70 & 0.74/0.69 & 0.75/0.70 & 0.71/0.62 & 0.71/0.68 \\
		\end{tabular}
	\end{center}
	\caption{ Topological measures for the five PNNs. Language names are abbreviated using their two-letter ISO language
	codes: EN (English), NL (Dutch), DE (German), ES (Spanish), and FR (French). $N$ is the number of nodes, $m$ is the number of edges,
	$\bar{k}$ is mean degree, \textbf{GC size} is the fraction of network nodes that are in the giant connected component, $C$ is clustering coefficient,
	$l$ is mean geodesic path length, $\alpha$ is the power law exponent of the degree distribution, and $r$ is the degree assortativity coefficient.
	Two values occurring in the table with a forward slash denote that quantity computed for the entire graph and only the giant component. Fits to degree
	distributions were performed via maximum likelihood~\cite{Clauset:2009} starting at $k=2$, except for French which began at $k=10$. Asterisks denote that
	the best fitting distribution is not strictly power law but rather truncated power law, as determined via a likelihood ratio test~\cite{Clauset:2009}.}
	\label{tab:topology}
\end{table}

\subsection{\label{sec:islands}Islands and Frequency Assortativity}

Given the relatively modest size of the giant connected component in all five languages (see Table~\ref{tab:topology}),
it is worth examining the connected component size (``island size'') distribution $P_c$ for each of the five PNNs. Power law distributions for the
sizes of the connected components $P_c$ have been previously observed in PNNs for both English and Spanish~\cite{Arbesman:2010b}.
Figure~\ref{fig:islands} shows that this power law distribution of component sizes is broadly shared over all five languages.
In fact, the island size distribution is more robustly power law than the PNN degree distribution itself, albeit over a
relatively modest range (less than a factor of 100).

We now remark on a previously unobserved feature of PNNs, again present in all five languages. All five languages show
a weak but statistically significant degree of word-frequency based assortativity. Simply, words of similar
usage frequency tend to be connected to each other in the PNN. We computed frequency assortativity by dividing the continuous word
frequency data into ten equal-mass bins and then computing an assortativity coefficient and jackknife standard deviation
using the definitions in Newman~\cite{Newman:2003}. The values ranged from $0.1$ in English to $0.24$ in Spanish, which
correspond to between 26 and 47 Jackknife standard deviations. This is weak relative to degree assortativity in these
networks (see Table~\ref{tab:topology}), but not insignificant on the scale of assortativity
coefficients found in other social, biological, and technological networks~\cite{Newman:2003}.

\subsection{\label{sec:lengthbias}DAS Graphs as Mixtures}

There is a deep physical basis for observing power laws in thermodynamics. Diverging length scales at critical points mean that there are correlations
at all scales in the system. Critical point behavior cannot depend on any quantity (like a force) with
an associated length scale, but rather only on scale-free quantities like symmetries and conservation laws. Critical point phenomena
then become universal, in the sense that the same behavior (critical exponents) is observed in systems that may have radically
different forces but the same set of symmetries.

The converse is not true. Observation of power laws does not necessarily indicate any
deep phenomena at work. Power laws in empirical data can arise from a wide variety of reasons, many of them mundane.
One of the simplest is Simon's famous demonstration~\cite{Simon:1955} that multiplicative (rather than additive)
random noise can yield heavy-tailed distributions. Another way to obtain power laws is via mixture distributions;
in this case apparent scale-free behavior arises by simply mixing several distributions, each with well-defined but different scales.

Indications that the degree distribution of the PNN for English results from a mixture of distributions of different
scales have been advanced by others~\cite{GP:2009}. Degree distributions for English PNNs separately constructed from short
and long (in phonemes) words showed different shapes and, at least for short words, displayed markedly less power-law behavior. In
Figure~\ref{fig:longshort} we show that this result also holds for the CLEARPOND English corpus, as well as for Dutch, German,
Spanish, and French. We divided all words in each corpus into two classes: monosyllabic and polysyllabic.

Figure~\ref{fig:longshort} clearly shows that connectivity among only monosyllabic words differs from polysyllabic word connectivity.  The monosyllabic
degree distributions look less like power laws than do the polysyllabic degree distributions, and monosyllabic words are in general more densely connected
than are polysyllabic words. This raises the possibility that the PNN degree distribution may arise as a mixture of distributions.
In all five languages, networks formed from polysyllabic words have degree distributions that are much closer to (truncated) power laws than are the
monosyllabic word networks.  In addition, note that (with the exception of French) the polysyllabic degree distributions are much more similar across the five
languages than the monosyllabic graph degree distributions or those of the full graphs (see Figure~\ref{fig:degreedist})

In Appendix~\ref{sec:syllables}, we look more closely at phonological neighbor graphs formed exclusively from monosyllabic or polysyllabic words,
and compare them to graphs containing all words in each corpus (see Table~\ref{tab:msps}). We found that some of the full PNN topological properties
are present in both the monosyllabic and polysyllabic networks (e.g., degree assortativity and clustering coefficient).
However, others are markedly different or disappear. The component or ``island'' size distribution $P_c$ is driven entirely by the polysyllabic words;
the monosyllabic words are almost completely connected (an unsurprising outcome of the DAS rule; shorter words, such as \textit{cat}, are much more likely to have DAS
neighbors than long words like \textit{catapult}). The full PNN graphs have short ($\sim 7$) average path lengths primarily because the monosyllabic
graphs have extremely short average path lengths ($\sim 5$) and the polysyllabic graphs have long ($\sim 10$) ones. When we compare the local properties of
the monosyllabic words in both the monosyllabic and full graphs, numbers of neighbors and second neighbors are highly
correlated. However, clustering is more weakly correlated, indicating that explanations of latencies in SWR that appeal to
node clustering~\cite{Chan:2009} coefficient as a predictor may be quite sensitive to whether or not polysyllabic words were included as items in the experiment.

At least three questions remain. First, do constraints imposed by the one-step neighbor DAS similarity measure explain the apparently universal topological
features seen across all five languages? If so, what explains the observed differences in the degree distributions in Figure~\ref{fig:degreedist}? Finally, how
much lexical structure is required to generate PNNs that resemble those of real languages?  In what follows, we address these three questions in detail.

\section{\label{sec:pseudolex}Pseudolexicons}

Figure~\ref{fig:longshort} and additional results that we present in the Appendix~\ref{sec:syllables} suggest that the truncated power law behavior observed in
the five PNNs might be the result of mixing subgraphs with different connectivity properties.  The left panel of Figure~\ref{fig:phonolen} again shows the
degree distributions for the five languages, this time with all homophones removed.  We discuss homophones in detail in Appendix~\ref{sec:lexical}; in brief,
we remove homophones because our random lexicon models produce phonological forms (rather than written words) directly and cannot properly account for homophones.
The right panel shows the distribution $P_l$ of words of length $l$ phonemes.  The $P_l$ distributions are underdispersed relative to Poisson (not shown); note
also that they are all zero-truncated, as there are no words in any language that consist of zero phonemes. A particularly intriguing feature of
the five language $P_l$ is that they cluster similarly to the degree distributions shown in Figure~\ref{fig:degreedist}. English and French together,
then German and Dutch, and Spanish by itself. While this could be entirely coincidental or a result of previously undetected cross-linguistic similarities,
below we will show that it is not.

\subsection{\label{sec:models} Models}

To determine which topological features of the PNNs arise due to specific features of real languages and which are driven purely by the
DAS connection rule, we adopt and extend an approach inspired by previous work on the English PNNs~\cite{GP:2009}.  We generate corpora of random phonological forms using
generative rules that include varying amounts of real linguistic detail.  We denote such a corpus of random strings of phonemes a pseudolexicon.  Each
pseudolexicon is paired with a target language, since all the models use some information from the real language for construction. Specifically, pseudolexicons are
created from the phonemic inventory of each language (the set of all phonemes that occur in the language), with lexicon size constrained to be approximately the same as the
real-language lexicon for the target language (for example, about 22,000 unique words [i.e., excluding homophones] for English CLEARPOND), and with the same form length
distribution as the target language. To match the length distribution, the length of each random string is first specified by drawing a random integer from a form length
distribution $P_l$ defined on the positive integers excluding zero. In all cases, the pseudolexicon has a form length distribution which we specify.  Specifically, we
consider the following six models for pseudolexicons.  We have named the models using terminology taken from the Potts~\cite{Potts:1952} and Ising~\cite{Ising:1925}
models.  Each includes progressively greater language-specific detail relating to phonological structure. We expected that we would get better a successively better
match to a given target-language PNN as we included more detail.

\begin{itemize}
	\item \textbf{Infinite Temperature (INFT).} Each phoneme in the string is drawn uniformly from the target language's phoneme inventory.
	\item \textbf{Noninteracting, Uniform Field (UNI).} Each phoneme in the string is drawn randomly using its observed frequency in the real language's lexicon.
	\item \textbf{Noninteracting, Consonant/Vowel Uniform Field (CVUNI).} Each position in the random string is either a consonant or a vowel drawn randomly using observed
	positional consonant/vowel frequencies in the real lexicon. Specifically, we use the real language's corpus to compute the position-dependent probability that position
	$l$ is a consonant or a vowel.  The particular consonant or vowel placed at that position is drawn uniformly from the respective set of items (lists of
	consonants and vowels).
	\item \textbf{Noninteracting, Consonant/Vowel Field (CV).}  Positions are selected to be consonants or vowels exactly as in CVUNI.  The particular consonant or
	vowel placed at each position is selected using observed frequencies of consonants and vowels from the real lexicon.
	\item \textbf{Noninteracting, Spatially Varying Field (SP).} Each phoneme is drawn randomly from real positional frequencies in the target lexicon.  For example,
	if a language has an inventory of twenty phonemes, we use the real lexicon to compute a $\pi_{l,x}$ that gives the probability that phoneme $x$
	occurs at position $l$, and then use this table to assign a phoneme to each position of the random string.
	\item \textbf{Nearest Neighbor Interactions (PAIR).} The first phoneme in each string is drawn using a positional probability.  Subsequent phonemes are drawn
	via the following rule.  If the phoneme at position $k$ is $x$, then the phoneme at position $k+1$ is drawn using the empirical probability (from the real
	lexicon) that phoneme $x'$ follows phoneme $x$.
\end{itemize}

We have listed the models in rough order of complexity; INFT uses the least amount of information about the real language's structure and PAIR the most.  We
note that while it is possible to generate real words (particularly short ones) from the models above, the vast majority of the strings produced bear no resemblance
to real words in any of the five languages.  The only model that avoids unpronounceable diphones is PAIR; in the other models unpronounceable
diphones occur frequently.

For each pseudolexicon, we discarded any duplicate items.  This is why we removed homophones from the real languages; we  did not generate orthographic tags for
the random phonological forms, so duplicated forms in the pseudolexicon all represent a single node.  We then formed a pseudo-PNN by using the DAS rule to connect
items in the pseudolexicon to one another.  As with the real PNNs, before any analysis we discarded nodes in the pseudo-PNNs with degree zero.  Figure~\ref{fig:fkpseudo}
shows the degree distribution of the Francis \& Kucera 1982 English corpus (FK)~\cite{FK:1982} and its six corresponding pseudolexicons.  We first show the fit to FK,
rather than CLEARPOND English, due to our ability to better control the contents of the FK corpus (see Appendix~\ref{sec:lexical}) for details).  Each of the six
pseudolexicons had as its input $P_l$ the empirical English $P_l$ (e.g.~Figure~\ref{fig:phonolen}, right panel).  We note that, while the sizes of the pseudolexicons
were fixed to the real-language target lexicon, once the pseudonetworks are formed, they may have fewer nodes than this, since many pseudowords may be neighborless and
hence not appear in the graph.

\subsection{\label{sec:english} English Networks}

Figure~\ref{fig:fkpseudo} shows that even minimal levels of linguistic realism yield a pseudo-PNN with a stikingly similar degree distribution to the real English PNN.
Even UNI, which includes nothing beyond overall phoneme frequencies and the empirical $P_l$, looks quite similar to FK.  Table~\ref{tab:fkpseudo}, which lists the same
topological properties we previously showed in Table~\ref{tab:topology} tells an even more compelling story.  First, the putatively lingustically relevant topological
properties discussed earier --- high clustering, short mean path length, high degree assortativity, and (to some extent) small giant components --- are present in all of
the pseudolexicons whose degree distributions match that of FK.  Giant component size is the least well-matched property in all of the models, though it is still smaller
than observed in many real-world networks.  Furthermore, even INFT, in which degree distribution (and hence mean degree) is a poor match to FK, has high clustering, short
mean path length, and high degree assortativity. INFT includes almost nothing about the target language except the form length distribution and the phoneme inventory.  We also
note the noisiness in the degree distribution of INFT.  While one might hypothesize that this is a result of its relatively small size, the degree distribution of INFT does
not become smooth even for larger (10,000 node) graphs (not shown).

\begin{table}[h]
	\begin{center}
		\renewcommand{\tabcolsep}{0.20cm}
		\begin{tabular}{c|ccccccc}
 			 & \textbf{FK} & \textbf{INFT} & \textbf{UNI} & \textbf{CVUNI} & \textbf{CV} & \textbf{SP} & \textbf{PAIR} \\
			\hline
			$N$ & 7861 & 1891 & 2922 & 1947 & 3022 & 3139 & 4346 \\
			$m$ & 22745 & 2841 & 7501 & 3532 & 8687 & 8811 & 12319 \\
			$\bar{k}$ & 5.79 & 3.0 & 5.13 & 3.63 & 5.74 & 5.61 & 5.67 \\
			\textbf{GC size} & 0.69 & 0.77 & 0.85 & 0.80 & 0.85 & 0.85 & 0.87 \\
			$C$  & 0.21 & 0.19 & 0.25 & 0.22 & 0.27 & 0.25 & 0.25 \\
			$l$ & 6.38 & 7.26 & 5.34 & 6.65 & 5.30 & 5.40 & 5.63 \\
			$\alpha$ & 1.0$^{*}$ & 1.0 & 1.0$^{*}$ & 1.0$^{*}$ & 1.0$^{*}$ & 1.0$^{*}$ & 1.0$^{*}$ \\
			$r$ & 0.67 & 0.60 & 0.45 & 0.64 & 0.48 & 0.49 & 0.45 \\
		\end{tabular}
	\end{center}
	\caption{ Topological measures for the FK English corpus and six pseudolexicons matched to it. All rows of the table are as described in
	Table~\ref{tab:topology}.}
	\label{tab:fkpseudo}
\end{table}

Figure~\ref{fig:cppseudo} and Table~\ref{tab:cppseudo} shows the same information for the CLEARPOND English database and pseudolexicons matched to it.  We note
first that all the conclusions that held for FK hold for CLEARPOND English.  Again, even a model as naive as UNI has a very similar degree distribution to the
real English PNN and very similar topological characteristics.  INFT, again despite having a degree distribution that is an extremely poor match to English CLEARPOND,
 has high clustering coefficient and high degree assortativity.  Compared to FK, some differences are evident.  Chiefly among them is that all the models now
 have too low of a mean degree, arising because the model degree distributions have large-$k$ tails that are too short.  However, given the analysis and discussion
 in Appendix~\ref{sec:lexical}, this is to be expected.  As discussed there, our models do not include analogs to inflected forms (e.g., WALK, WALKS, WALKED). We
also have not attempted to model homophones (which have been removed in our pseudo-lexicon PNNs) or proper nouns. All three of these item types preferentially affect
the tail shape of $P_k$.  We also note that the CLEARPOND-matched pseudolexicons tend (except for SP) to undershoot the English giant component size, though they still
match the fundamental observation that the GC is a relatively small portion of the full network.

\begin{table}[h]
	\begin{center}
		\renewcommand{\tabcolsep}{0.20cm}
		\begin{tabular}{c|ccccccc}
 			 & \textbf{EN} & \textbf{INFT} & \textbf{UNI} & \textbf{CVUNI} & \textbf{CV} & \textbf{SP} & \textbf{PAIR} \\
			\hline
			$N$ & 18252 & 3192 & 5911 & 3942 & 6219 & 7098 & 8705 \\
			$m$ & 59965 & 3748 & 13281 & 6857 & 16373 & 18821 & 20922 \\
			$\bar{k}$ & 6.6 & 2.35 & 4.49 & 3.48 & 5.27 & 5.31 & 4.81 \\
			\textbf{GC size} & 0.65 & 0.36 & 0.41 & 0.47 & 0.38 & 0.63 & 0.35 \\
			$C$ & 0.21 & 0.19 & 0.25 & 0.22 & 0.27 & 0.25 & 0.25 \\
			$l$ & 6.81 & 16.7 & 6.53 & 9.43 & 5.77 & 10.7 & 9.35 \\
			$\alpha$ & 1.0$^{*}$ & 1.0$^{*}$ & 1.0$^{*}$ & 1.0$^{*}$ & 1.0$^{*}$ & 1.0$^{*}$ & 1.0$^{*}$ \\
			$r$ & 0.70 & 0.83 & 0.74 & 0.85 & 0.68 & 0.71 & 0.72 \\
		\end{tabular}
	\end{center}
	\caption{ Topological measures for the CLEARPOND English corpus (EN) and six pseudolexicons matched to it. All rows of the table are as described in
	Table~\ref{tab:topology}.}
	\label{tab:cppseudo}
\end{table}

\subsection{\label{sec:fivelang} Five Language Pseudonetworks}

We now compare pseudo-PNNs to real PNNs for all five languages: English, Spanish, Dutch, German, French.  For this comparison, we used only the UNI model,
since it has a very similar degree distribution to the English PNN $P_k$ despite containing almost no information about real language phonology and constraints.  In
each case, the pseudo-PNN is matched in total corpus size and form length distribution to its target language.  The left panel of Figure~\ref{fig:fivepseudo}
shows the true degree distributions for the five language PNNs (shown also in Figure~\ref{fig:phonolen}) and the right panel of Figure~\ref{fig:fivepseudo} shows
the pseudo-PNNs using the UNI model.  Furthermore, Table~\ref{tab:fivepseudo} shows topological parameters for Spanish, French, German, and Dutch and their matched UNI
pseudo-PNNs.  We omit English in Table~\ref{tab:fivepseudo} because that information is contained in Table~\ref{tab:cppseudo}.

Figure~\ref{fig:fivepseudo} and Table~\ref{tab:fivepseudo} together show that, as in English, the UNI model is able to come remarkably close in shape and topological
properties to the real phonological neighbor networks, despite not resembling the real language's phonology in any way.  The clustering of the five language degree
distributions for the pseudo-PNNs mimics that seen in the real PNNs, particularly in the manner in which Spanish is separated from the other
languages. Given the way the UNI pseudo-PNNs were constructed, this grouping must be driven entirely by the form length distribution.  Table~\ref{tab:fivepseudo}
shows that the pseudo-PNNs match their target languages quite well overall, with some properties extremely similar, e.g.~clustering coefficients and degree assortativity.

\begin{table}[h]
	\begin{center}
		\renewcommand{\tabcolsep}{0.20cm}
		\begin{tabular}{c|cccccccc}
 			 & \textbf{FR} & \textbf{pFR} & \textbf{ES} & \textbf{pES} & \textbf{DE} & \textbf{pDE} & \textbf{NL} & \textbf{pNL} \\
			\hline
			$N$ & 12164 & 7854 & 20018 & 2198 & 16787 & 4141 & 14943 & 3938 \\
			$m$ & 32753 & 21577 & 31812 & 2852 & 35402 & 8749 & 31697 & 9408 \\
			$\bar{k}$ & 5.38 & 5.49 & 3.16 & 2.60 & 4.17 & 4.23 & 4.24 & 4.79 \\
			\textbf{GC size} & 0.72 & 0.36 & 0.43 & 0.32 & 0.57 & 0.33 & 0.55 & 0.57 \\
			$C$ & 0.28 & 0.27 & 0.19 & 0.19 & 0.21 & 0.25 & 0.16 & 0.27 \\
			$l$ & 7.13 & 7.49 & 9.49 & 9.99 & 8.88 & 5.34 & 8.5 & 11.59\\
			$\alpha$ & 1.0$^{*}$ & 1.0$^{*}$ & 1.0$^{*}$ & 1.0$^{*}$ & 1.0$^{*}$ & 1.0$^{*}$ & 1.0$^{*}$ & 1.0$^{*}$ \\
			$r$ & 0.59 & 0.75 & 0.70 & 0.65 & 0.71 & 0.67 & 0.73 & 0.66 \\
		\end{tabular}
	\end{center}
	\caption{ Topological measures for four phonological neighbor networks (FR, ES, DE, NL) and matched UNI pseudo-PNNs (pFR, pES, pDE, pNL).
	All rows of the table are as described in Table~\ref{tab:topology}.}
	\label{tab:fivepseudo}
\end{table}

\begin{figure}[ht!]
	\begin{center}
		\includegraphics[width=\columnwidth]{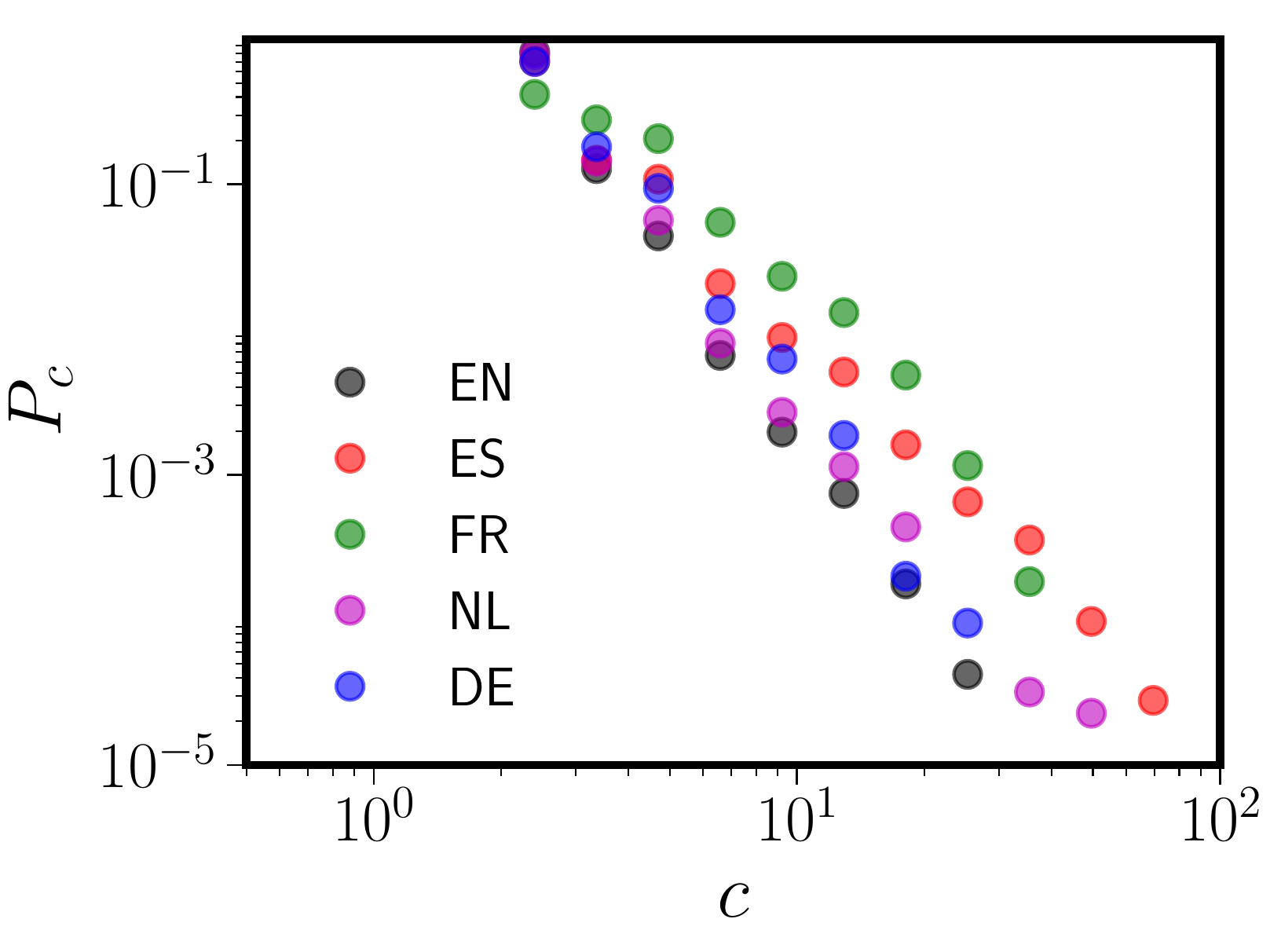}
	\end{center}
	\caption{Size distributions $P_c$ vs. $c$ for the connnected components (``islands'') in all five languages. Languages
	are abbreviated using their two-letter ISO codes (see Table~\ref{tab:topology}). The giant component has been excluded
	from this figure for all five languages; it sits far to the right for each language. The minimum island size is two
	because we have removed any singleton nodes (loners) from the PNNs, as discussed in Section~\ref{sec:data}.}
	\label{fig:islands}
\end{figure}

\begin{figure*}[htp!]
	\begin{center}
		\subfigure{
			\includegraphics[width=0.48\columnwidth]{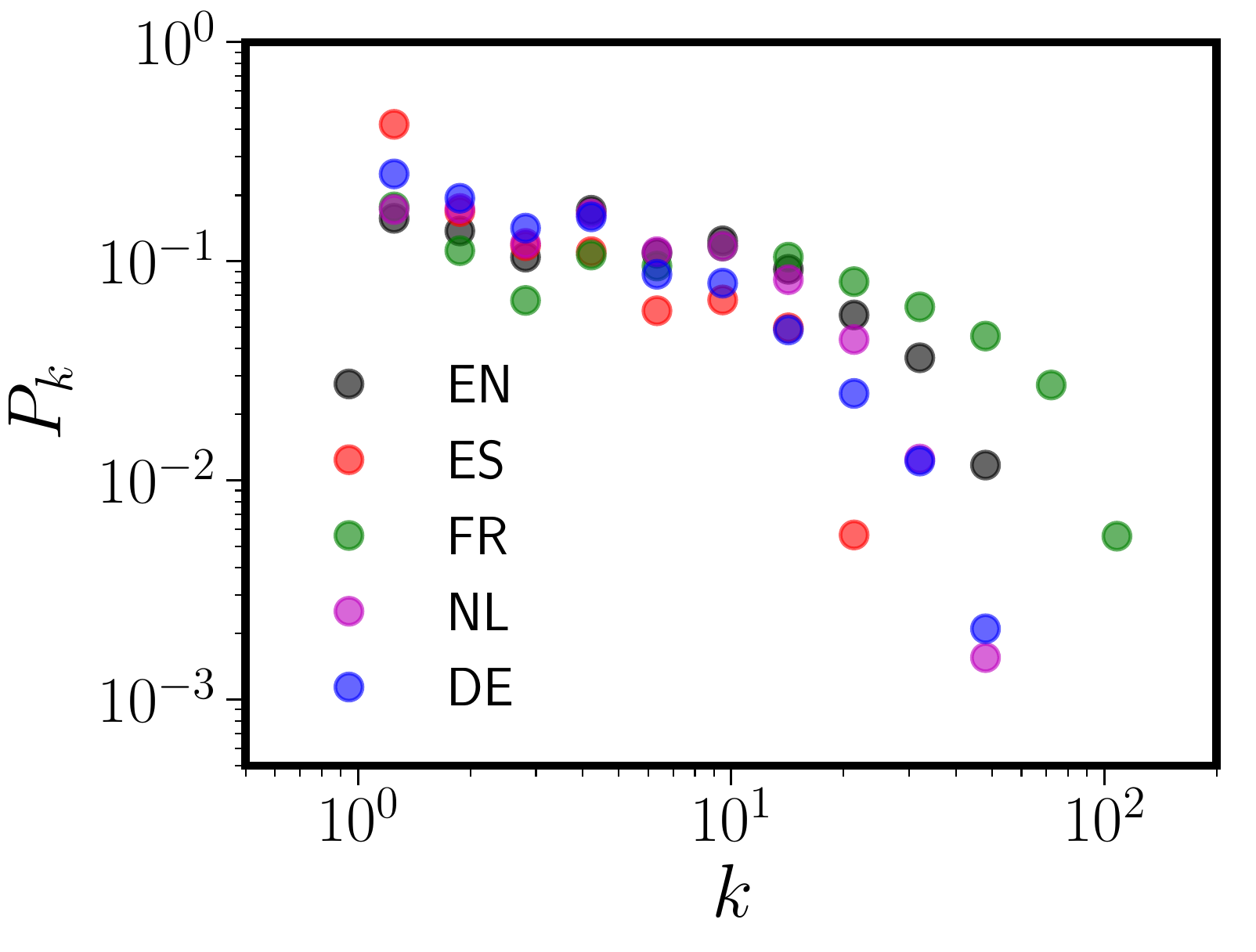}
			}
		\subfigure{
			\includegraphics[width=0.48\columnwidth]{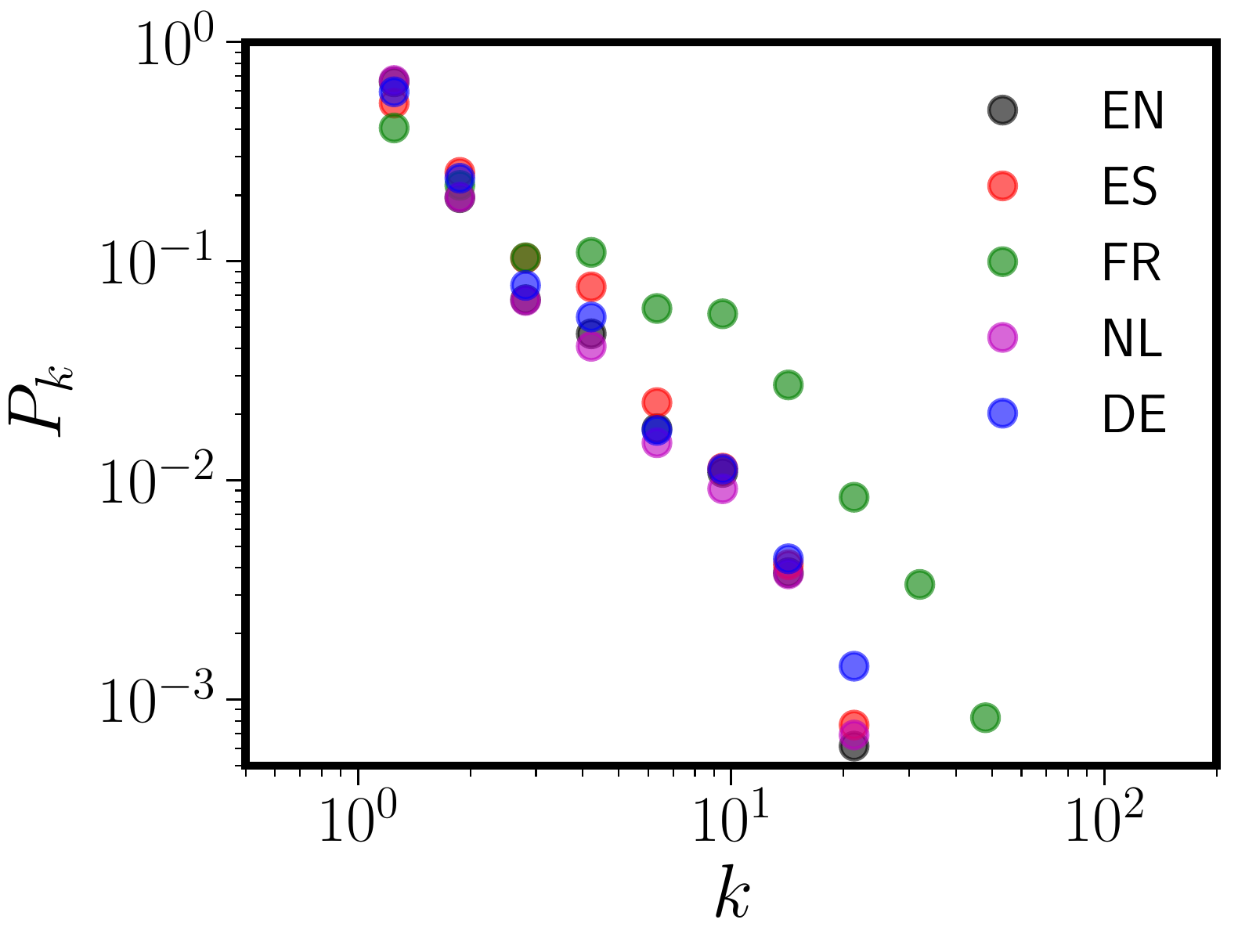}
		}
	\end{center}
	\caption{Degree distributions $P_k$ vs. $k$ for PNNs formed from exclusively monosyllabic (left panel) or polysyllabic (right panel)
	words in each lexicon. Each language is abbreviated by its two letter ISO code; see the caption to Table~\ref{tab:topology} for the
	key to these codes.}
	\label{fig:longshort}
\end{figure*}

In Figure~\ref{fig:islands} we showed that the component size distributions for all five language PNNs follow a power law, even moreso than the degree distributions
for the PNNs themselves. This has previously only been observed in English and Spanish~\cite{Arbesman:2010b}.  However, even these component size distributions
do not arise out of any fundamental or universal phonological properties.  In the left panel of Figure~\ref{fig:pseudoislands} we reprint Figure~\ref{fig:islands} to
allow easy comparisons.  In the right panel we show component size distributions for the five pseudo-PNNs.  While the span of $P_c$ is somewhat reduced in the pseudo-networks,
all the pseudographs clearly have power law size distributions with exponents similar to their target languages.  Thus, even the island size distribution is
essentially an artifact of the neighbor definition.

\subsection{\label{sec:sens} Sensitivity to the Form Length Distribution}

The previous section demonstrates that the topological properties of phonological neighbor networks constructed using the one-step DAS rule
are driven not by any real linguistic feature but by the connection rule itself.  While the resulting PNNs are remarkably insensitive to the
degree to which real phonological constraints are used in their construction, we have also shown that the PNNs \textit{are} sensitive to the shape
of the form length distribution $P_l$.  In this section, we investigate that sensitivity further.  We do that by generating four more UNI lexicons with
different input form length distributions and compare the resulting pseudo-PNNs to the FK English database.  The four form length distributions are
as follows.

\begin{itemize}
	\item \textbf{EMP}. $P_l$ is the empirical form length distribution for FK, exactly as in Figure~\ref{fig:fkpseudo} and Table~\ref{tab:fkpseudo}.
	\item \textbf{ZTP(1x)}. $P_l$ is a zero-truncated Poisson (ZTP) model fit to the empirical distribution.  The ZTP distribution has the form
		\begin{equation}
			P_{\mathrm{ZTP}}(k;\lambda) = \frac{\lambda^k}{(e^{\lambda} - 1)k!},
		\end{equation}
		which assuming independence among the empirical length values $x_i$ leads to a model likelihood
		\begin{equation}
			L(\lambda) = \prod_{i=1}^N \frac{\lambda^{l_i}}{(e^{\lambda} - 1)l_i!}
		\end{equation}
		in which $\lambda$ can be determined via numerical maximization $L(\lambda)$.
	\item \textbf{ZTP(1.5x)} This model is a zero-truncated Poisson model for $P_l$ with a mean equal to 1.5 times the mean of the ML $\lambda$ of ZTP-1X.
	\item \textbf{GEO}.  Here $P_l$ follows a geometric distribution
		\begin{equation}
			P_{\mathrm{GEO}}(k;p) = p(1-p)^{k-1}
		\end{equation}
		for which the parameter $p$ is chosen to make the mean of GEO equal to the mean of the empirical English $P_l$.
\end{itemize}

\begin{figure*}[htp!]
	\begin{center}
		\subfigure{
			\includegraphics[width=0.48\columnwidth]{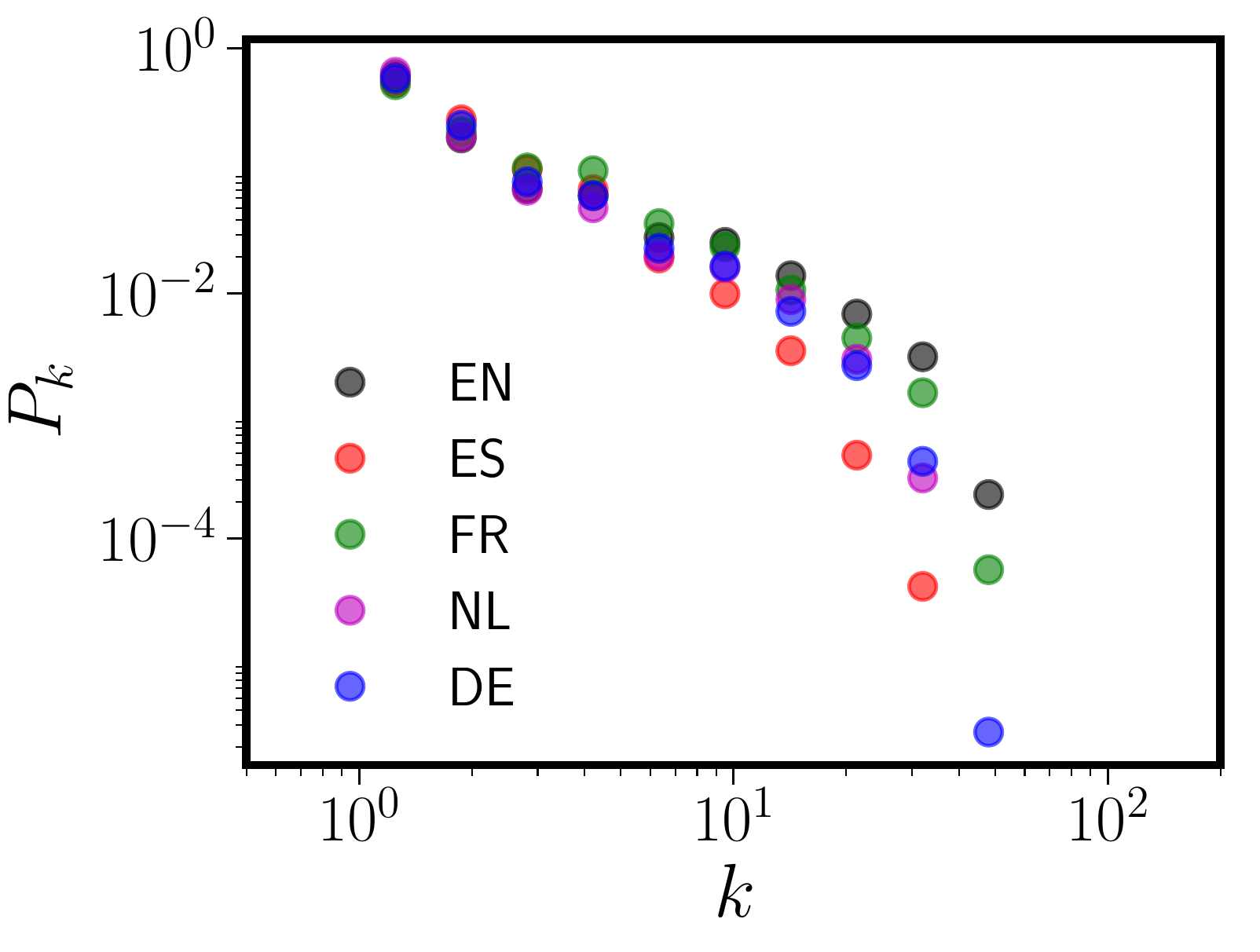}
			}
		\subfigure{
			\includegraphics[width=0.465\columnwidth]{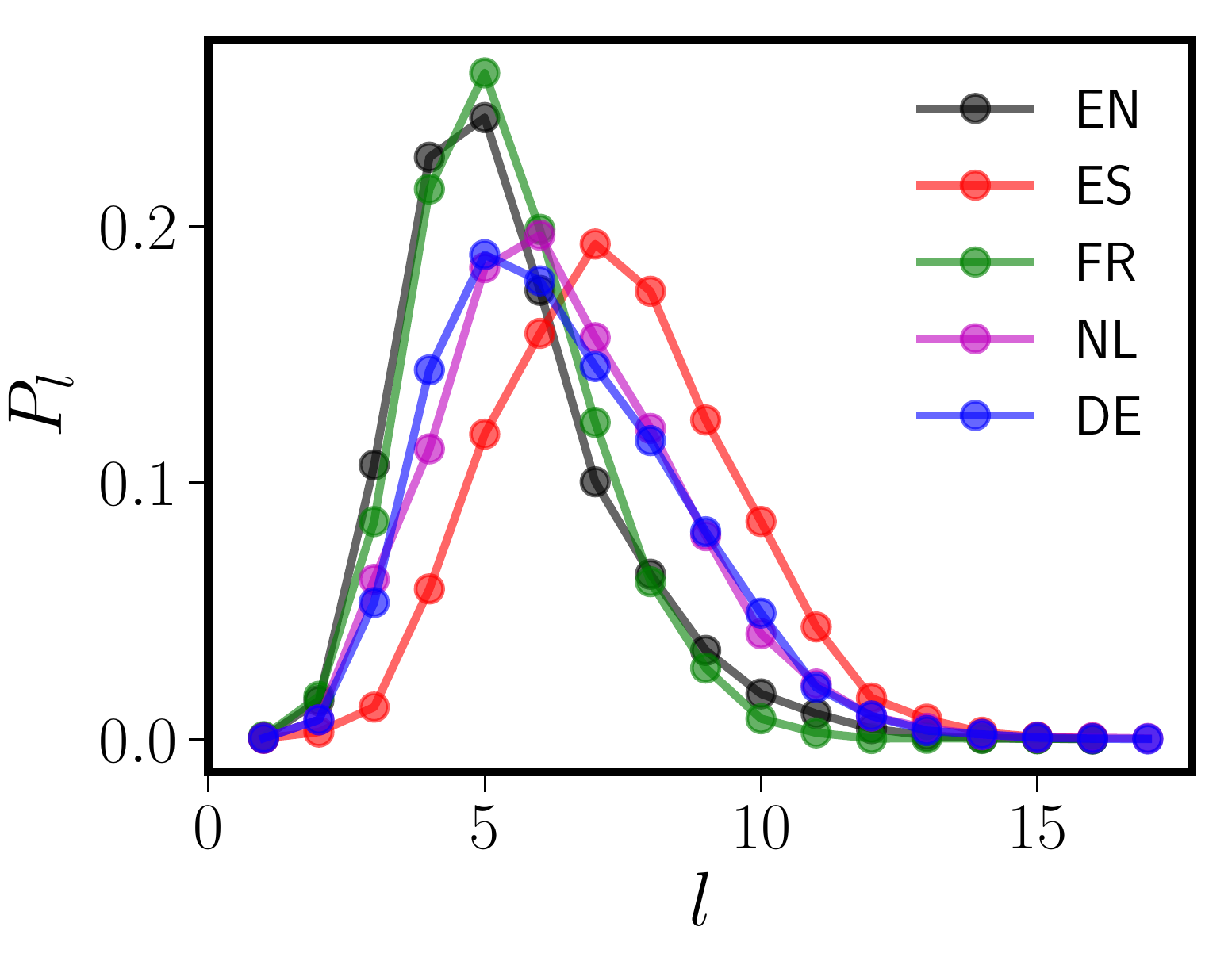}
		}
	\end{center}
	\caption{The left panel shows the degree distributions $P_k$ versus $k$ for the five CLEARPOND PNNs.  Compare to Figure~\ref{fig:degreedist}; this
	figure differs because homophones have been removed from the graphs as detailed in Appendix~\ref{sec:lexical}.  The right panel shows the distribution
	$P_l$ of phonological form lengths in each of the five languages from the CLEARPOND corpora. Note that all these distributions are only defined for
	$l \geq 1$; length zero words do not exist.}
	\label{fig:phonolen}
\end{figure*}

\begin{figure}[htp!]
	\begin{center}
			\includegraphics[width=\columnwidth]{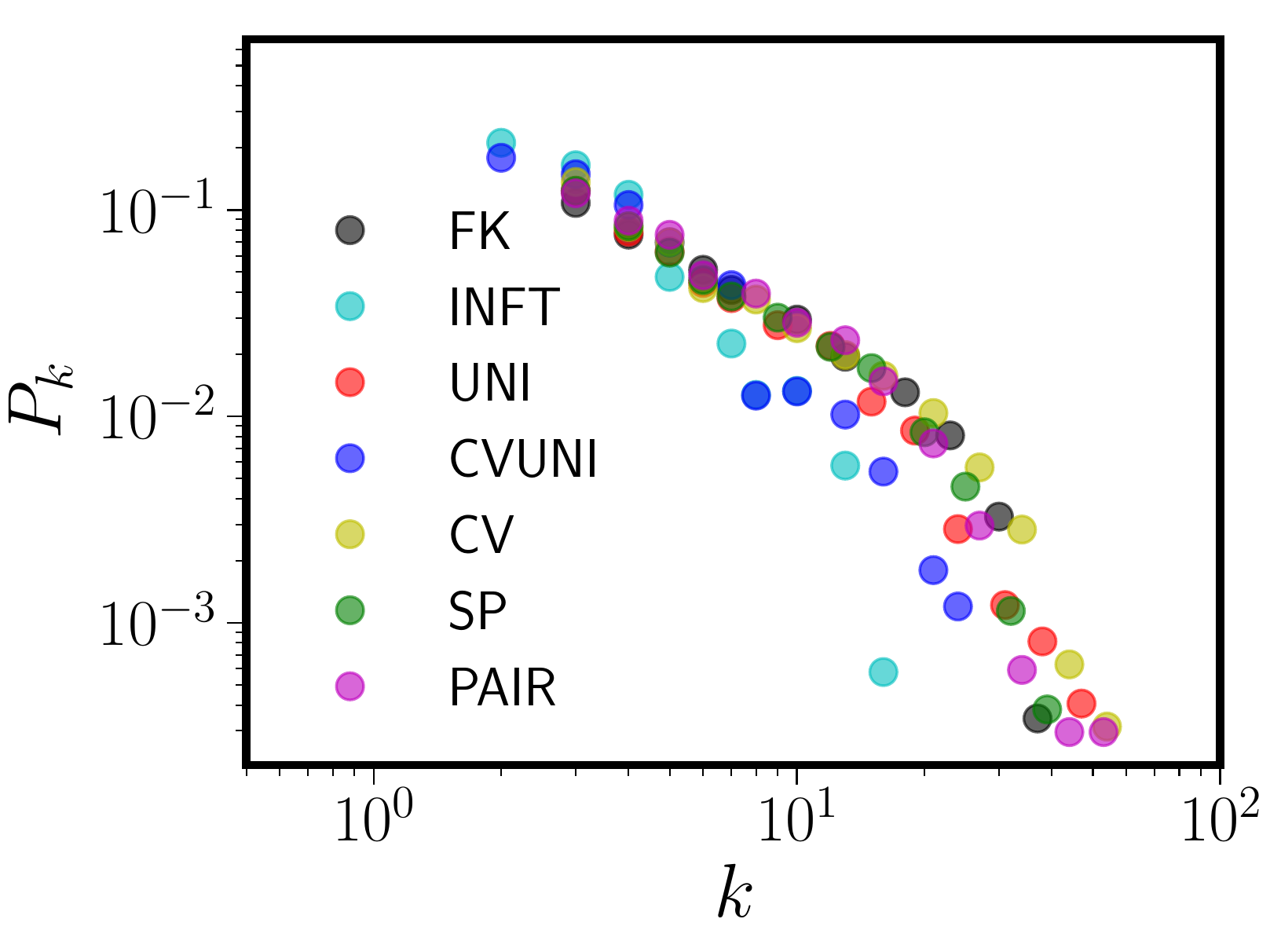}
	\end{center}
	\caption{Degree distributions $P_k$ versus degree $k$ for the Francis and Kucera 1982 corpus (FK) along with the six pseudolexicons fit to it. See the text
	for a key to the abbreviations for the pseudolexicons.}
	\label{fig:fkpseudo}
\end{figure}

\begin{figure}[htp!]
	\begin{center}
			\includegraphics[width=\columnwidth]{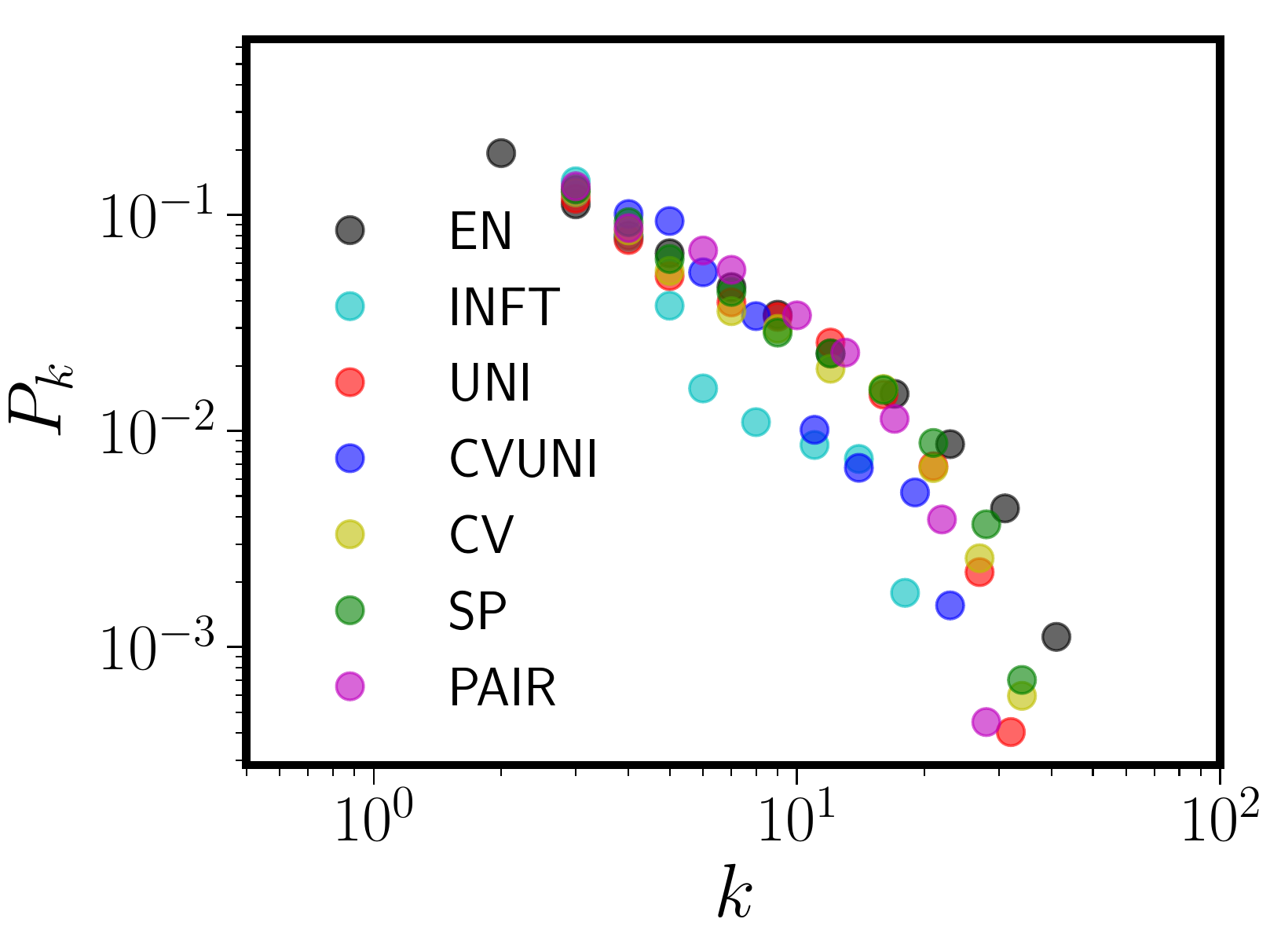}
	\end{center}
	\caption{Degree distributions $P_k$ versus degree $k$ for the CLEARPOND English corpus (EN) along with the six pseudolexicons fit to it.  See the text for
	a key to the abbreviations for the pseudolexicons.}
	\label{fig:cppseudo}
\end{figure}

\begin{figure*}[htp!]
	\begin{center}
		\subfigure{
			\includegraphics[width=0.48\columnwidth]{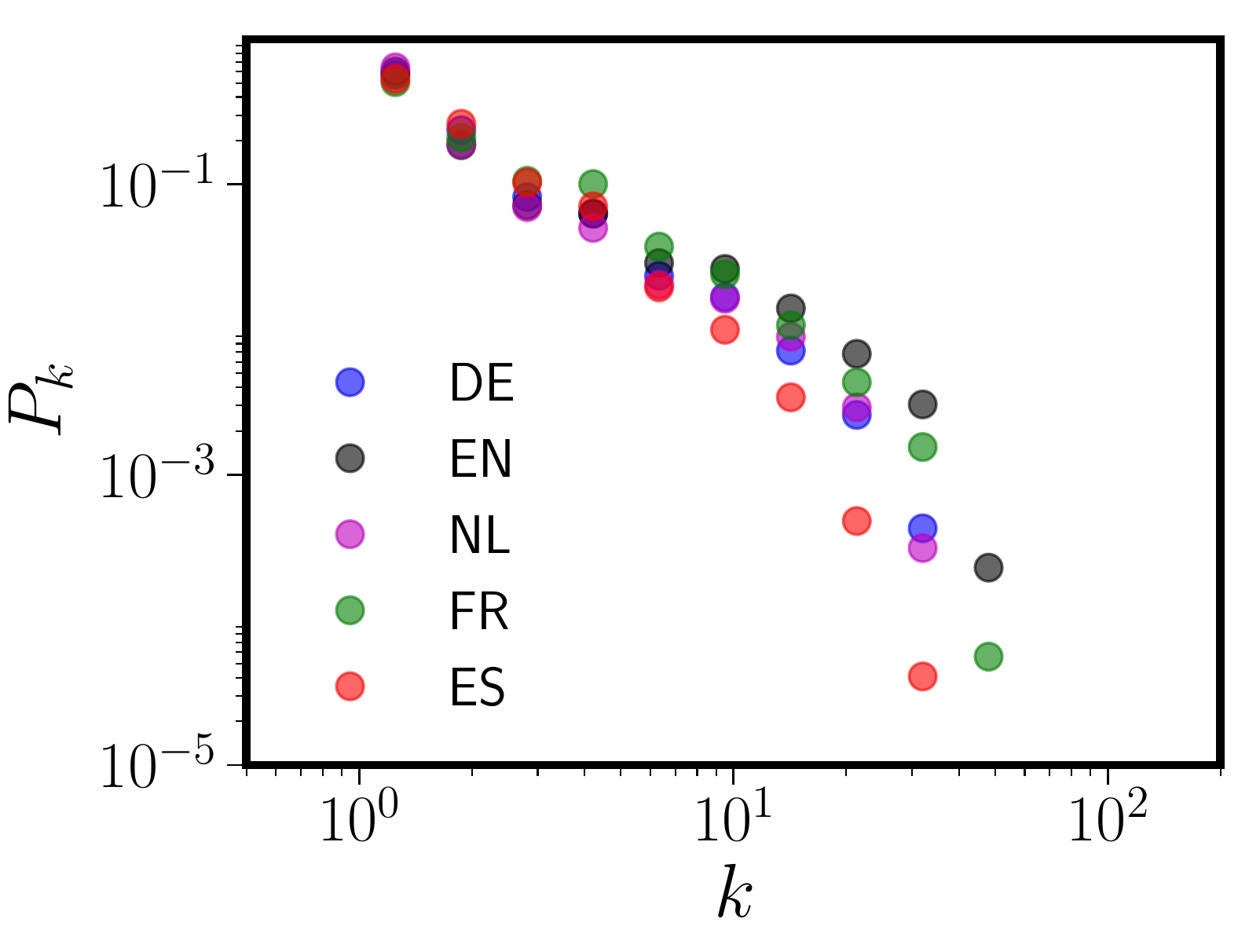}
			}
		\subfigure{
			\includegraphics[width=0.48\columnwidth]{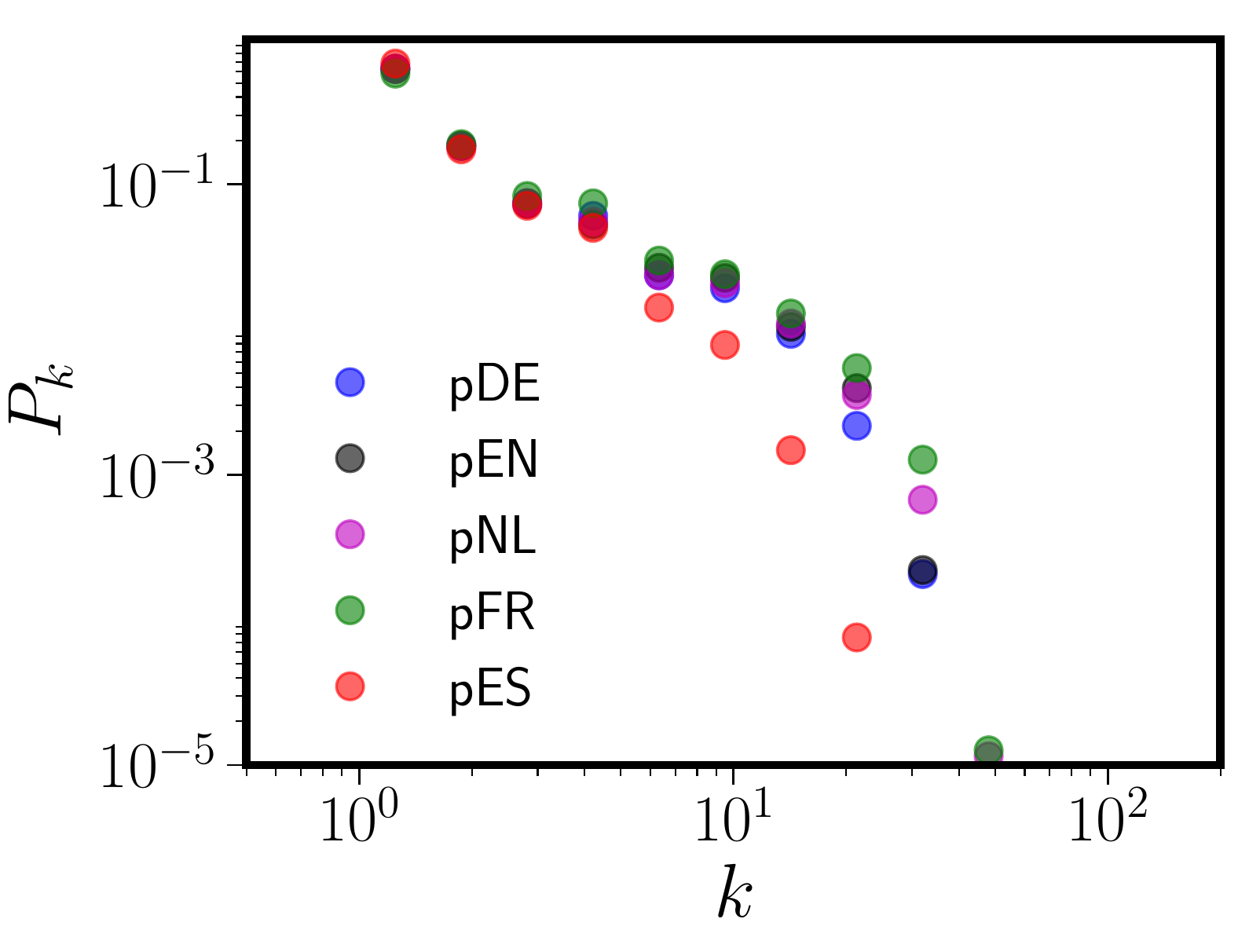}
		}
	\end{center}
	\caption{The left panel shows the degree distributions $P_k$ versus $k$ for the five CLEARPOND PNNs.  This is a reprint of the left panel of
	Figure~\ref{fig:phonolen}, so that comparisons may be more easily made.  The right panel shows degree distributions for pseudo-PNNs, each of which
	is produced using the UNI model (see Section~\ref{sec:models}) and matched to the target language.}
	\label{fig:fivepseudo}
\end{figure*}

\begin{figure*}[htp!]
	\begin{center}
		\subfigure{
			\includegraphics[width=0.48\columnwidth]{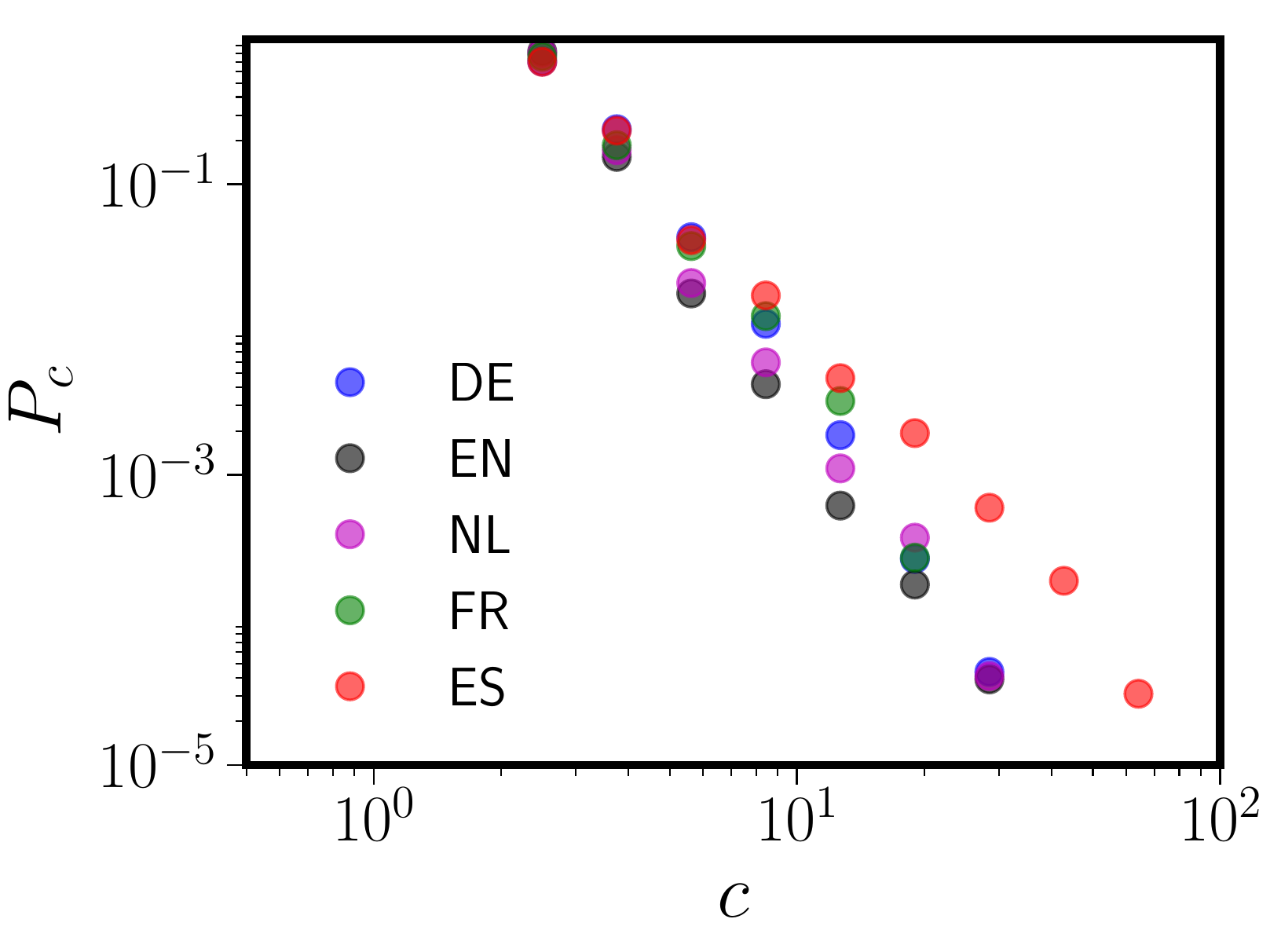}
			}
		\subfigure{
			\includegraphics[width=0.48\columnwidth]{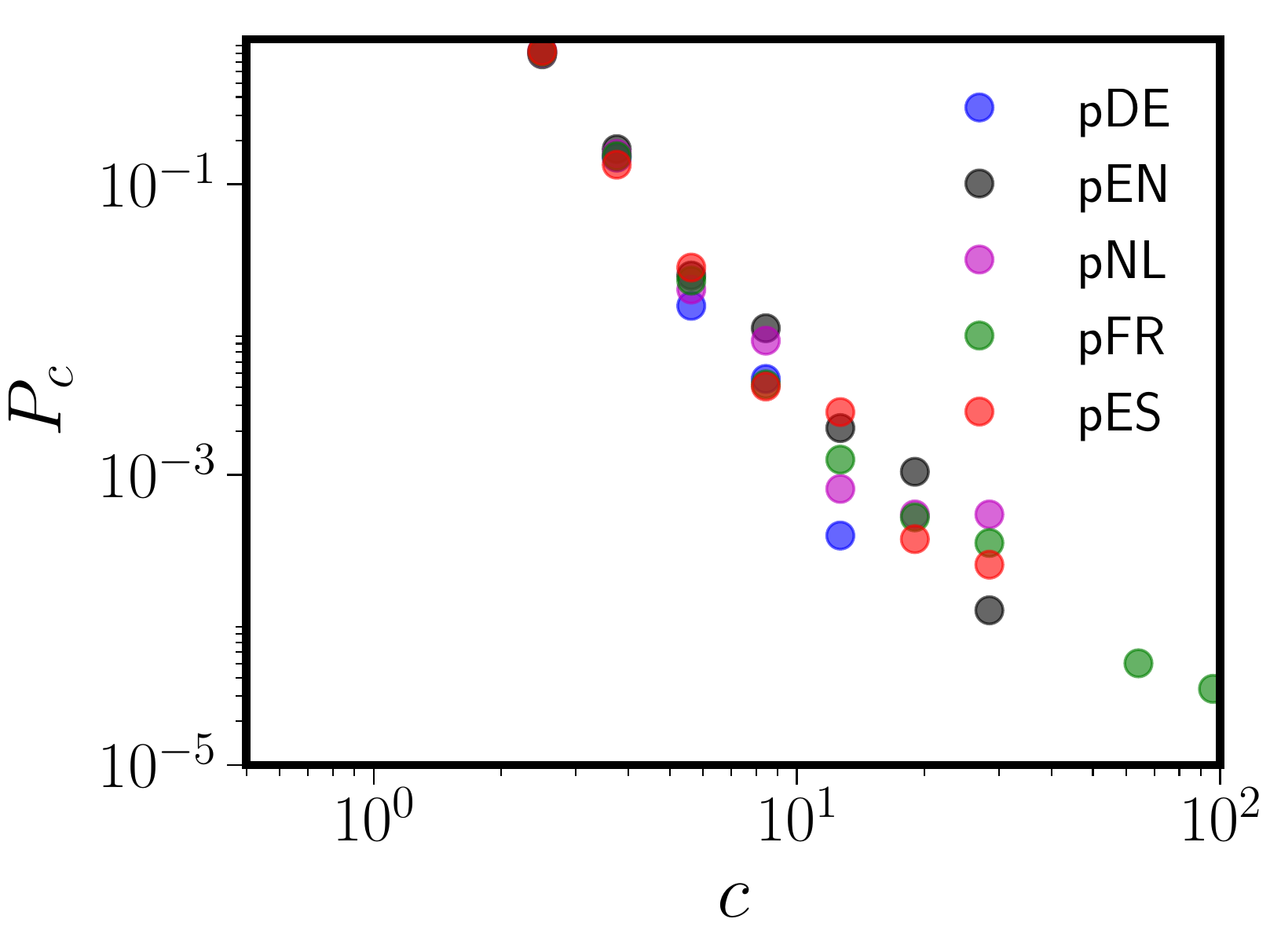}
		}
	\end{center}
	\caption{The left panel shows the component size distribution $P_c$ versus $c$ (compare to Figure~\ref{fig:islands}). The right panel shows
	component size distributions for pseudo-PNNs, each of which is produced using the UNI model (see Section~\ref{sec:models}) and matched to the target language.}
	\label{fig:pseudoislands}
\end{figure*}

\begin{figure}[htp!]
	\begin{center}
			\includegraphics[width=\columnwidth]{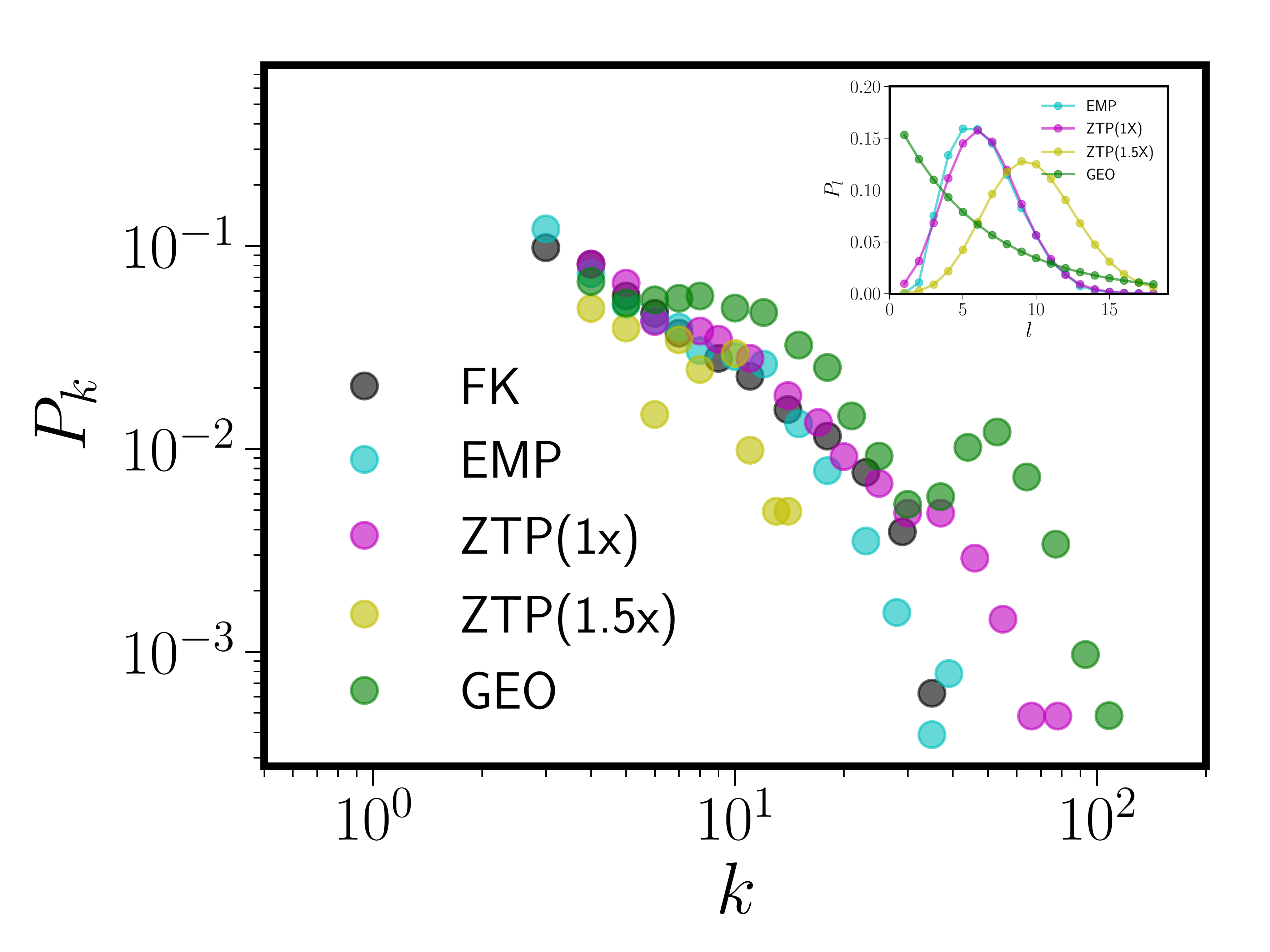}
	\end{center}
	\caption{Degree distributions (main panel) for UNI pseudo-PNNs constructed using four different phonological length distributions: the empirical English
  form length distribution (EMP), a zero-truncated Poisson fit to the empirical distribution (ZTP(1x)), a zero-truncated Poisson with shifted mean (ZTP(1.5x)),
  and a geometric distribution (GEO) with the same mean as EMP.  The real FK network is shown for comparison, and the inset shows the four different form
  length distributions.}
	\label{fig:plsens}
\end{figure}

We chose ZTP(1x) as a simple but relatively poor approximation to the form length distributions of the real languages; the real form length distributions are all
underdispersed relative to Poisson.  ZTP(1.5x) as compared to ZTP(1x) is similar to the difference between the Spanish $P_l$ and those of English and French
(see Figure~\ref{fig:phonolen}). GEO is chosen to have an identical average length to English phoneme strings, but otherwise has a shape radically different than any $P_l$
we observe. Figure~\ref{fig:plsens} shows degree distributions for pseudo-PNN constructed using UNI pseudolexicons, with each of these four choices for $P_l$,
along with the degree distribution of FK for comparison.  The inset in Figure~\ref{fig:plsens} shows the form length distribution used to produce the pseudolexicon
yielding the PNN in the main panel.  Table~\ref{tab:plsens} compares the topological properties of those four pseudo-PNNs to FK and each other.

\begin{table}[h!]
	\begin{center}
		\renewcommand{\tabcolsep}{0.20cm}
		\begin{tabular}{c|ccccc}
 			 & \textbf{FK} & \textbf{EMP} & \textbf{ZTP(1x)} & \textbf{ZTP(1.5x)} & \textbf{GEO} \\
			\hline
			$N$ & 7861 & 2959 & 2753 & 211 & 3592 \\
			$m$ & 22745 & 7954 & 13127 & 304 & 35938 \\
			$\bar{k}$ & 5.79 & 5.38 & 9.54 & 2.88 & 20.0 \\
			\textbf{GC size} & 0.69 & 0.85 & 0.85 & 0.64 & 0.95 \\
			$C$ & 0.21 & 0.24 & 0.28 & 0.19 & 0.35  \\
			$l$ & 6.38 & 5.19 & 4.48 & 4.71 & 3.73 \\
			$\alpha$ & 1.0$^{*}$ & 1.0$^{*}$ & 1.0$^{*}$ & 1.74 & 1.0$^{*}$ \\
			$r$ & 0.67 & 0.44 & 0.53 & 0.46 & 0.49 \\
		\end{tabular}
	\end{center}
	\caption{Topological measures for four UNI pseudo-PNNs (EMP, ZTP-1X, ZTP-1.5X, GEO) and the real FK phonological neighbor network. All rows of the table
	are as described in Table~\ref{tab:topology}.}
	\label{tab:plsens}
\end{table}

It is clear from Figure~\ref{fig:plsens} that the shape of the PNN degree distribution is extremely sensitive to the form length distribution.  Even the
relatively small differences in the shape of EMP and ZTP(1x) lead to large changes in the tail mass of the degree distribution.  The difference between the
degree distribution of ZTP(1x) and ZTP(1.5x) is similar to the difference between the degree distributions of English or French and Spanish (see Figure~\ref{fig:phonolen}).
In addition, Table~\ref{tab:plsens} shows that the PNN made from ZTP(1.5x) is much smaller (fewer nodes and edges) than any of the other models. This is expected
given the reduction in probability of short phonological forms in ZTP(1.5x) when compared to EMP, ZTP(1x), or GEO; the probability that two strings from the UNI
pseudolexicon that differ in length by one unit or less are neighbors decays exponentially with string length.  Note also from Table~\ref{tab:plsens} that no matter
what effect $P_l$ has on the degree distribution of the resulting PNN, all graphs show high clustering coefficients, short mean free paths, and high degree assortativity.

\section{\label{sec:conclusion}Conclusion}
We have shown that observed ``universal'' topological features of phonological neighbor networks~\cite{Arbesman:2010a}
--- truncated exponential degree distributions, high clustering coefficients, short mean free paths, high degree assortativity and small
giant components --- are string rather than language universals.  That is, inferences from networks based on similarity regarding language ontogeny or phylogeny are
suspect, in light of our analyses demonstrating that similar network structures emerge from nearly content-free parameters. One might object to this strong interpretation.
The DAS rule obviously captures important relations that predict significant variance in lexical processing due to similarity of phonological forms in the lexicon.
Networks based on DAS are able to extend DAS's reach, as was previously demonstrated with clustering coefficient~\cite{Chan:2009,Chan:2010}. Note, though, that clustering
coefficient relates to familiar concepts in word recognition that have not been deeply explored in the spoken domain: the notion of neighbors that are friends or enemies at
specific positions, discussed by McClelland and Rumelhart in their seminal work on visual word recognition~\cite{McClelland:1981}. Consider a written word like \textit{make},
with neighbors such as \textit{take}, \textit{mike}, and \textit{mate}. \textit{Take} is an enemy of the first letter position in \textit{make}, but a friend at all other
letter positions, where it has the same letters. A written word with a clustering coefficient approaching 1.0 would have many neighbors that all mismatch at the same position
(thus making them neighbors of each other). A word with a similar number of neighbors but a low clustering coefficient  (approaching $N/L$, that is, $N$ neighbors evenly
distributed of $L$ [length] positions) would have more evenly distributed neighbors. For spoken word recognition, the results of Chan and Vitevitch~\cite{Chan:2009} suggest
that a high clustering coefficient exacerbates competition because it is heavily loaded on a subset of phoneme positions, creating high uncertainty. In our view, this
reveals important details about phonological competition, but not ontogeny or phylogeny of English, or other specifically linguistic structure. Indeed, given the similarity
in the distribution of clustering coefficients (among other parameters) in English and in our abstract PNNs, we interpret instances of (e.g.) high clustering coefficient
as string universals rather than language universals.

While phonological neighbor network topology is largely insensitive to the degree of real phonological structure in the lexicon used to construct the
neighbor network, we found some amount of sensitivity to the input form length distribution $P_l$.  Even relatively subtle changes in $P_l$ can lead
to observable changes in the degree distributions of the resulting neighbor networks, and differences among the five languages we studied here can
be almost wholly attributed to differences in form length distributions among the five languages.  However, even this sensitivity is only partial.
Form length distributions that look nothing like any of the languages we consider here (GEO, although GEO may partially resemble the $P_l$ of a language
like Chinese), that generate network degree distributions that we do not observe, still yield high clustering coefficients, short mean free paths, and
high degree assortativity.  The question of what leads to a given language's $P_l$ is a question about language evolution that will be much more
difficult to explain, though some parallels might be drawn with work that seeks to understand the evolution of orthography~\cite{Christiansen:2003,Plotkin:2000,
FerreriCancho:2003}.

At an even deeper level, it may be perilous to attach too much meaning to the topology of any similarity network of phonological forms,
at least with respect to human performance in psycholinguistic tasks. This is because these networks do not ``do'' anything; they have no
function. They are not connectionist networks that attempt to model phoneme perception, like TRACE~\cite{McClelland:1986} or TISK~\cite{Hannagan:2013}.
No matter how they are constructed, they are basically static summaries of the structure of the speech lexicon; they do not perform a
processing function.  Insofar as the similarity measure aligns with latency data from human spoken words tasks (e.g.~picture naming~\cite{Chan:2010},
lexical decision~\cite{Chan:2009}, etc.), network properties may encode some features of human performance.  While there is evidence that some aspects of
human task performance may be predicted from features of neighbor networks~\cite{Vitevitch:2008,Iyengar:2012,Chan:2009,Chan:2010,Siew:2017}, it is clear from our study
that care must be taken in interpreting the results of studies of phonological networks. If the static structure of the lexicon were to be paired with a
dynamics that represents mental processing, it would be possible to test the utility of phonological similarity networks for explaining human performance
in psycholinguistic tasks.

\section*{Acknowledgements}

We acknowledge the seed grant program from the Connecticut Institute for Brain and Cognitive Sciences (CT IBaCS) for supporting this work.

\appendix
\section{\label{sec:syllables} Syllable Level Analysis}

To deterimine the number of syllables, we count vowels and dipthongs in the phonological transcription of each word.  In addition, we
correct for words that end in a phonological `l' with no vowel preceeding the final phoneme. For example, the English word
\textit{able} has only a single vowel but is a two-syllable word.  We note here that syllable \textit{boundaries} are much harder to determine, but we do not need to
decompose the word into its constituent syllables for any of our analysis.

For each language we built two additional graphs: one for the monosyllabic (MS) words only and one for the polysyllabic (PS) words
only. Just as in the full PNN, these two new graphs used the DAS rule to determine if two words should be connected by an edge. We
show results for English and Dutch are in Table~\ref{tab:msps}. Table~\ref{tab:msps} shows that the topological properties of the PNNs
arise by mixing two very different kinds of graphs. For quantities like the clustering coefficient and degree assortativity, this mixing
is very mild. The MS graphs tend to cluster more strongly than the PS graphs, and vice versa for degree assortativity, but the differences
are not extreme. This is not the case for the rest of the topological measures. Despite having far fewer nodes, the MS graphs have tenfold
greater edge density. The MS graphs are almost completely connected; all ``islands'' in the English and Dutch PNNs are induced by the structure
of PS graphs. Mean geodesic paths are quite short in the MS graphs and long in the PS graphs. The MS graphs do not have power law degree
distributions at all; that arises due to mixing with the PS graphs (all power-law or truncated power law) in the full graph.

\begin{table}[h]
	\begin{center}
		\begin{tabularx}{\textwidth}{c|cccccc}
			& \textbf{EN MS+PS} & \textbf{EN MS} & \textbf{EN PS} & \textbf{NL MS+PS} & \textbf{NL MS} & \textbf{NL PS}\\
			\hline
			$N$ & 18983 & 5979 & 13004 & 15630 & 2808 & 12552 \\
			$m$ & 76092 & 50232 & 19808 & 36158 & 16785 & 18396 \\
			$d$ & 0.0004 & 0.003 & 0.0002 & 0.0003 & 0.004 & 0.0002\\
			$\bar{k}$ & 8.01 & 16.8 & 3.0 & 4.71 & 11.96 & 2.93 \\
			\textbf{GC size} & 0.66 & 0.98 & 0.46 & 0.31 & 0.97 & 0.43\\
			$C$ & 0.23/0.28 & 0.3/0.3 & 0.19/0.26 & 0.16/0.23 & 0.31/0.30 & 0.13/0.20 \\
			$l$ & 6.68 & 4.63 & 10.3 & 4.62 & 11.8 & 8.73 \\
			$\alpha$ & 1.0$^{*}$ & - & 1.04$^{*}$ & 1.84$^{*}$ & - & 1.72 \\
			$r$  & 0.73/0.70 & 0.65/0.65 & 0.74/0.66 & 0.74/0.69 & 0.59/0.59 & 0.74/0.65 \\
			$r_f$ & 0.104(4)$\; [26\sigma]$ & 0.068(4)$\; [15\sigma]$ & 0.089(7)$\; [13\sigma]$ & 0.126(5)$\; [25\sigma]$ & 0.055(8)$\; [7\sigma]$ & 0.100(7)$\; [14\sigma]$\\
		\end{tabularx}
	\end{center}
	\caption{Topological measures for graphs produced from the CLEARPOND English and Dutch corpora. MS+PS is the
	full  PNN (see also Table~\ref{tab:topology}), MS is a graph formed from only the monosyllabic
	words, and PS a graph formed from only the polysyllabic words. With the exception of edge density $d$
	and frequency assortativity coefficient $r_f$, all symbols in this table are the same as those in
	Table~\ref{tab:topology}, and the quantities in the tabhle separated by forward slashes have the same meaning
	as in Table~\ref{tab:topology}. Edge density is defined as $2m/N(N-1)$, where $m$ is the number of edges
	and $N$ the number of nodes in the graph.}
	\label{tab:msps}
\end{table}

\begin{figure}[htp!]
	\begin{center}
			\includegraphics[width=\columnwidth]{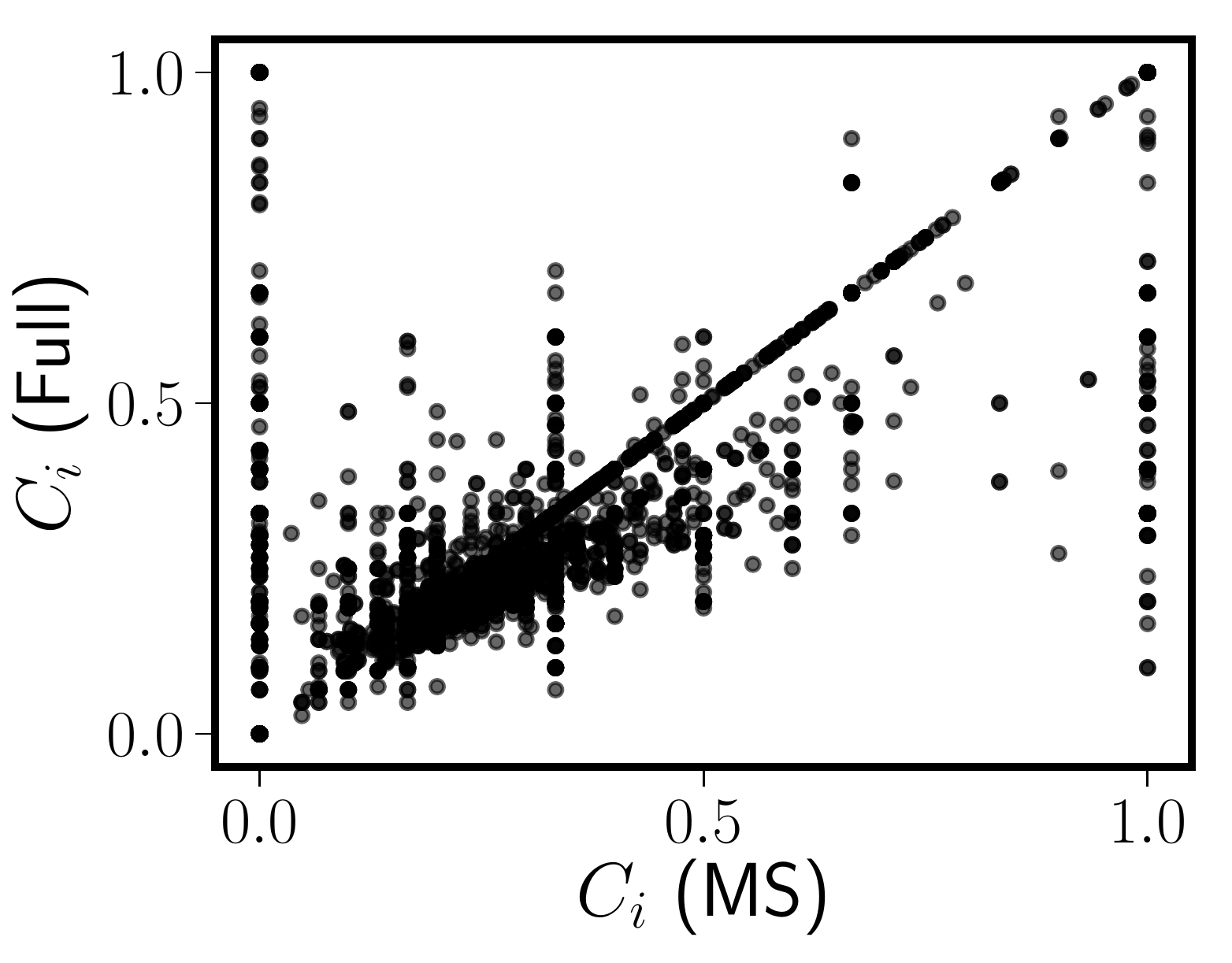}
	\end{center}
	\caption{Clustering coefficient for each MS node in the MS graph ($x$-axis) and in the full English CLEARPOND PNN ($y$-axis). The $R^2$
	of the correlation between the two sets of values is $0.8$.}
	\label{fig:clustmsps}
\end{figure}

We also compared node-level topology for the MS words in the MS only graph and the full PNN (MS+PS). Most
quantities are almost perfectly correlated for these two: these include number of neighbors (degree), number of second
neighbors, and eigenvector centrality. All of these quantites are highly correlated with $R^2 \geq 0.95$.   Node clustering coefficient
for the MS words in the two English graphs is more weakly similar ($R^2 = 0.8$), with large outliers (see Figure~\ref{fig:clustmsps}).
It would be interesting to revisit the proposed relationship between node clustering and spoken word recognition~\cite{Chan:2009} facility
in light of these findings.

When we performed the same syllable-level calculations for the other three languages in the CLEARPOND database, we
find a consistent story (results not shown). In all cases, MS giant component sizes are much larger than PS GC sizes,
MS edge densities are close to tenfold larger, and MS mean geodesic path lenghts are much shorter. PS degree distributions
are well-fit by truncated power laws and have much more consistent power law exponents than we see for the full
PNNs in Table~\ref{tab:topology}. All five languages except Dutch have power law exponents in the range $1.0-1.04$. Dutch
is better fit by a non-truncated power law with exponent $1.72$ (see Table~\ref{tab:msps}). Furthermore, clustering coefficient
and degree assortativity are similar in the MS and PS graphs, just as in English and Dutch. As in English and Dutch, clustering
coefficients are larger for the MS graphs in German, Spanish, and French.

\section{\label{sec:lexical}Lexical Issues}

In this section we discuss some lexical issues that are relevant to the pseudolexicon models we construct in Section~\ref{sec:pseudolex}.
Table~\ref{tab:worddegree} shows the thirteen most highly connected words in English CLEARPOND.  What should be clear from Table~\ref{tab:worddegree}
is that proper nouns (\textit{Lowe}) and homophones (\textit{see},\textit{sea}) are overrepresented.  Our pseudolexicon models directly generate phonological
forms with no orthographic tags; we thus cannot represent homophones in our models and must remove them from the PNN graphs for comparison.  Another
category of words not represented in Table~\ref{tab:worddegree} that we cannot easily model is inflected forms, for example word-final `s' for plurals
in English.  Lemmas and their inflected forms occur much more frequently than expected at random, particular for longer (multisyllabic) words.  We
discuss the effects of these three categories of words (homophones, inflected forms, and proper nouns) on the resulting PNN degree distribution, and explain
our protocol for their removal.

\begin{table}[h]
	\begin{center}
		\begin{tabularx}{2in}{Y|Y}
 			\textbf{word} & \textbf{degree} \\ \hline
			Lea & 68 \\
			Lee & 68 \\
			Lew & 66 \\
			loo & 66 \\
			lieu & 66 \\
			Lai & 63 \\
			lye & 63 \\
			lie & 63 \\
			Lowe & 62 \\
			low & 60 \\
			male & 60 \\
			see & 60 \\
			sea & 60 \\
		\end{tabularx}
	\end{center}
	\caption{The thirteen words in English CLEARPOND with the highest degree.  Note the prevalence in this list of (\textit{i}) proper
	nouns and (\textit{ii}) homophones (e.g.~,\textit{see},\textit{sea}).}
	\label{tab:worddegree}
\end{table}

We began with the Francis \& Kucera 1982 English corpus~\cite{FK:1982} (hereafter FK), which consists of the Brown corpus of English words, along
with prononuciation (phonological transcription) for every item and an indication as to whether the item is a lemma or an inflected form.  We used FK rather than
CLEARPOND here because, as shown below, we lack some of this word-level information in CLEARPOND and cannot remove all three categories of words in the CLEARPOND
PNNs.    We successively removed inflected forms, proper nouns, and homophones from FK as follows.
\begin{itemize}
	\item \textbf{Proper Nouns}. Any word whose orthographic (written) form begins with a capital letter is assumed to be a proper noun.  This rule applies equally well
	to the PNNs for FK and CLEARPOND.  However, we emphasize here that because of the rules for capitalization in German (all nouns are capitalized), we cannot systematically
	remove proper nouns for all five languages in CLEARPOND.
	\item \textbf{Inflected Forms}. FK includes lemma numbers for all the words, so we can simply remove any words that are not lemmas.  We do not have this information
	for any words in CLEARPOND and thus cannot remove them.  To try to remove inflected forms in CLEARPOND we could, for example, remove all words with word-final
	phonological `z'. This would remove English plurals but also improperly remove some lemmas (\textit{size}).  Even if this were desirable, we would need different rules for
	all five languages.  Therefore we are forced to keep all inflected forms in the CLEARPOND PNNs.
	\item \textbf{Homophones} Homophones are items with identical phonological transcriptions but different orthography.  These are relatively simple to remove in both
	FK and CLEARPOND English, and the same procedure works in any language.  We search the nodes for sets of items with identical phonological transcriptions.  For
	example, \textit{see} and \textit{sea} would comprise one homophone set in English, and \textit{lieu}, \textit{loo}, and \textit{Lou} another.  One of the items from
  each homophone set, chosen at random, is kept in the PNN and the nodes corresponding to all other items in the set are deleted.
\end{itemize}

\begin{figure}[htp!]
	\begin{center}
			\includegraphics[width=\columnwidth]{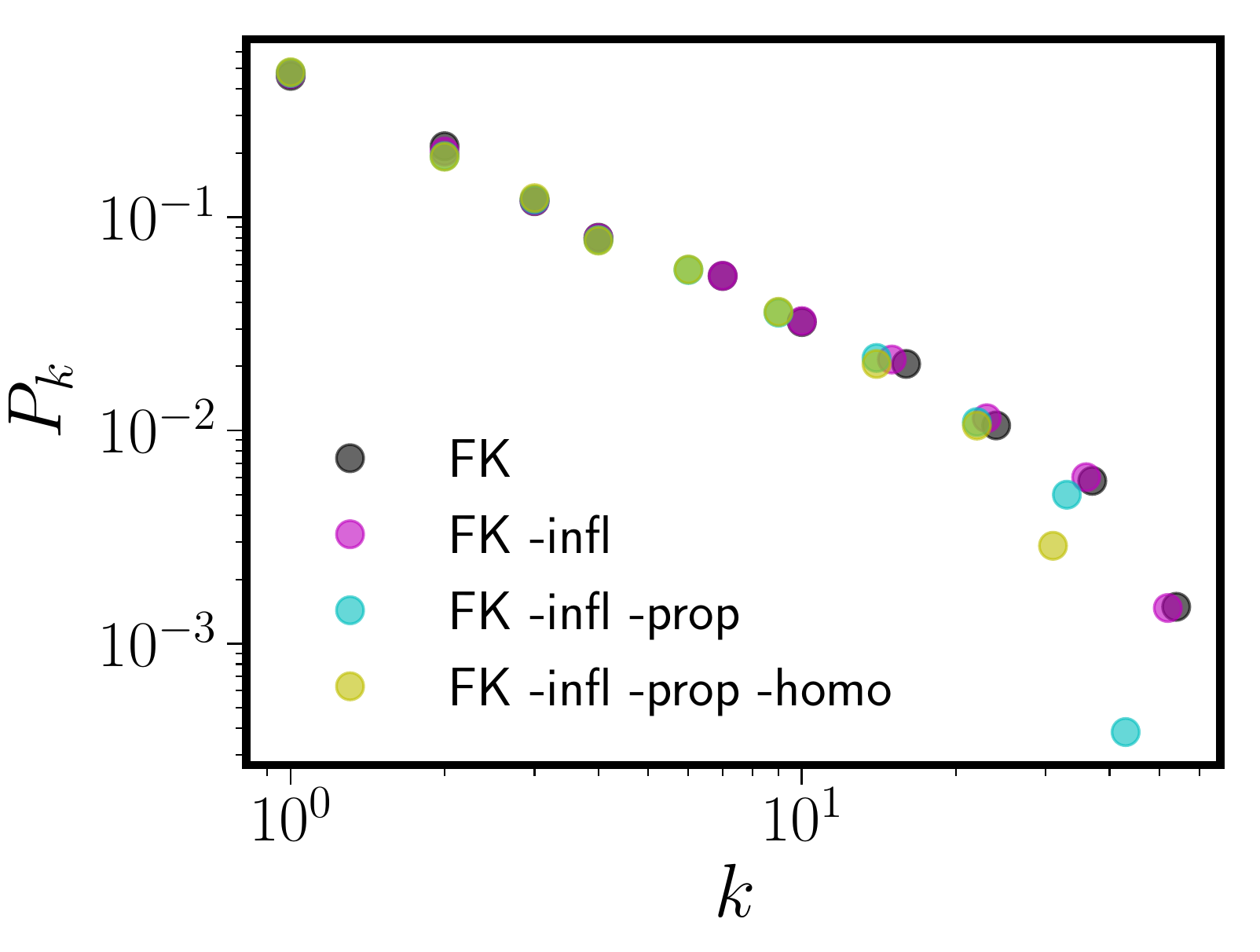}
	\end{center}
	\caption{Double log plot of the degree distribution $P_k$ against $k$ for the FK phonological neighbor network (black circles) when inflected forms (magenta),
	proper nouns (cyan), and homophones (yellow) are successively removed.}
	\label{fig:FKremoval}
\end{figure}

Figure~\ref{fig:FKremoval} shows the degree distribution of the FK PNN when inflected forms, proper nouns, and homphones were successively removed.  Two features
of this figure deserve mention.  First, the main effect of these classes of words is in the tail of the degree distribution.  Secondly, removal of inflected forms
causes very little change compared to removal of proper nouns and homophones.  It is relatively easy to understand why the largest changes to the degree distribution
occur at large $k$, at least for homophones.  Consider a single orthographic form $w$ that is also a homophone with degree $d$.  All of the other orthographic forms in
its homophone set are connected to both $w$ and all of the $d$ neighbors of $w$.  If there are $N$ words in the homophone set, we end up with $N$ nodes each with degree
$d$ + $N-1$.  Thus, homophone sets can boost the degree of both their neighbors (since a neighbor of one is a neighbor of all other words in the set) and the homophones
themselves.  As an example, a homophone set of size 10 in which one of the words has 10 neighbors yields 10 nodes with degree 19.  Removing members of the homophone set
will therefore tend remove nodes of large degree and therefore shift the tail of $P_k$.

\begin{figure}[htp!]
	\begin{center}
			\includegraphics[width=\columnwidth]{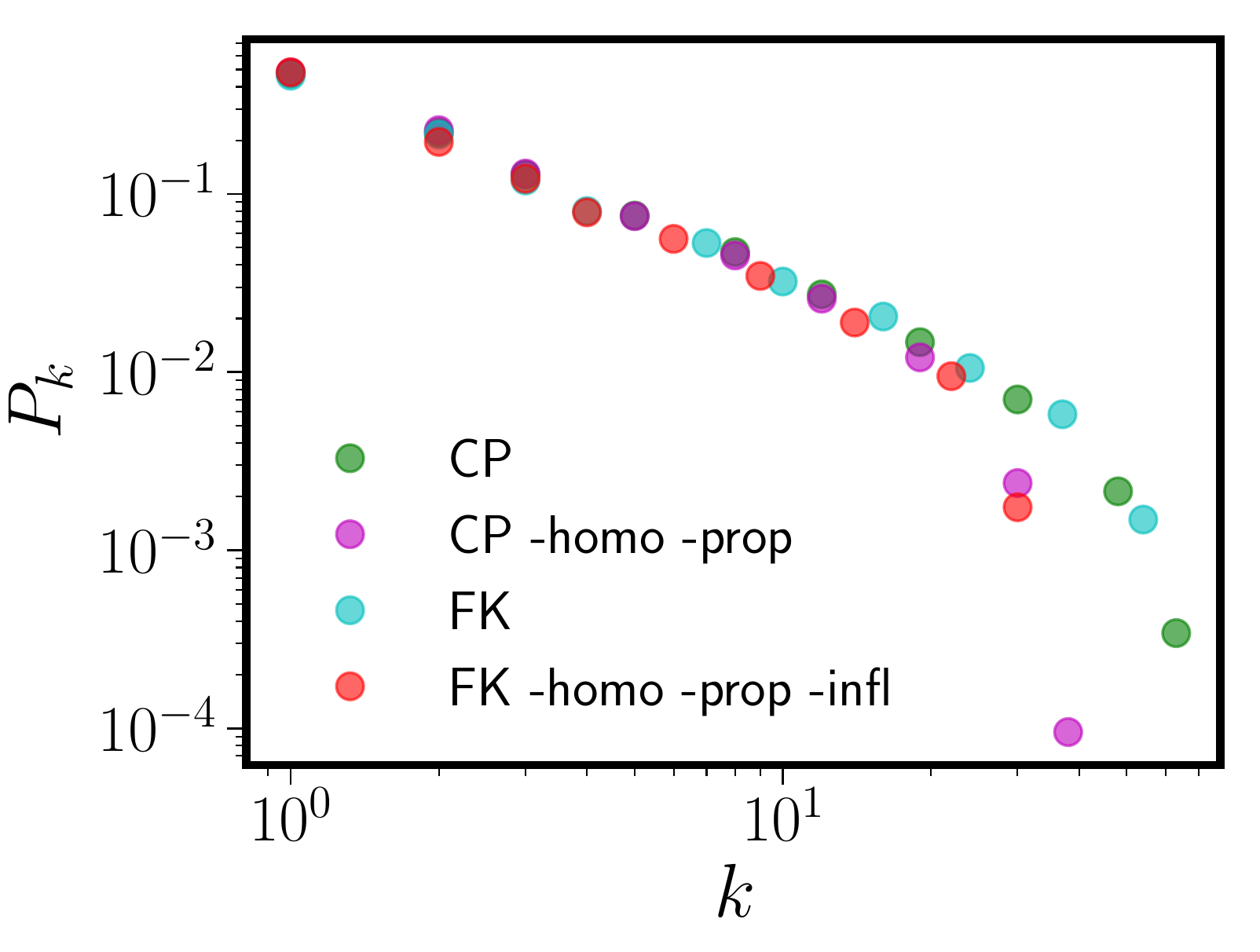}
	\end{center}
	\caption{Double log plot of the degree distribution $P_k$ against $k$ for the CLEARPOND English PNN and FK when various classes of words are removed.  The CLEARPOND
	English PNN degree distribution is show unaltered (green) and after homophones and proper nouns are removed (magenta), while FK is show unaltered (cyan) and when
	homophones, proper nouns, and inflected forms are all removed (red).}
	\label{fig:CPremoval}
\end{figure}

Figure~\ref{fig:CPremoval} compares removal in English CLEARPOND to FK.  We first note that, despite being based on completely different corpora, the unaltered English
CLEARPOND and FK yield similar PNNs.  In addition, as in FK, removal of homophones and proper nouns in CLEARPOND tends to truncate the tail of
the degree distribution.  As we noted above, the only class of words that we can consistenly remove from all five CLEARPOND languages is homophones, and we remove these
for all model comparisons.  The number of homophone sets, mean set size, and the number of nodes removed from the graph when the removal procedure described
above is implemented, for each of the five languages in CLEARPOND is shown in Table~\ref{tab:homophones}.  Table~\ref{tab:homophones} indicates that for all languages
except French, the majority of homophone sets are pairs like \textit{see} and \textit{sea}.

\begin{table}[h]
	\begin{center}
		\renewcommand{\tabcolsep}{0.40cm}
		\begin{tabular}{c|ccc}
 			\textbf{Language} & $N_H$ & $\mu_H$ & \textbf{Nodes Removed} \\
			\hline
			EN & 731 & 2.09 & 795  \\
			DE & 440 & 2.10 & 485  \\
			ES & 1059 & 2.03 & 1123 \\
			FR & 9013 & 2.63 & 14735\\
			NL & 417 & 2.08 & 449 \\
		\end{tabular}
	\end{center}
	\caption{Number of homophone sets $N_H$, mean homophone set size $\mu_H$ and the number of nodes removed from the CLEARPOND PNNs when homophones are removed.
	Note the wide variation in the number of homophones across the five languages.}
	\label{tab:homophones}
\end{table}

\bibliography{short,fivelang}

\clearpage

\clearpage

\clearpage

\clearpage

\clearpage

\clearpage

\clearpage

\clearpage

\clearpage

\clearpage

\clearpage

\clearpage

\end{document}